\documentclass[prd,aps,floats,amssymb,superscriptaddress,nofootinbib,altaffilletter,twocolumn]{revtex4}

\def\be{\begin{equation}}
\def\ee{\end{equation}}
\def\bea{\begin{eqnarray}}
\def\eea{\end{eqnarray}}
\def\pa{\partial}

\def\fn{\footnote}

\def\case#1/#2{\textstyle\frac{#1}{#2}}

\begin{document}
\title{VARIATIONS ON THE SEVENTH ROUTE TO RELATIVITY}
\setcounter{footnote}{2}
\author{Edward Anderson}
\email[]{eda@maths.qmul.ac.uk}
\addtocounter{footnote}{4}
\affiliation{{\em  Astronomy Unit, School of Mathematical Sciences, Queen Mary, London E1 4NS, U.K. }}
\date{10 Febuary 2003}
\begin{abstract}

Wheeler asked how one might derive the Einstein--Hamilton--Jacobi equation 
from plausible first principles without any use of the Einstein field equations 
themselves.  In addition to Hojman, Kucha\v{r} and Teitelboim's `seventh route to relativity' 
partial answer to this, there is now a `3-space' partial answer due to Barbour, Foster and \'{O} Murchadha (BF\'{O}) 
which principally differs in that general covariance is no longer presupposed.
BF\'{O}'s formulation of the 3-space approach is based on \it best-matched \normalfont actions like the lapse-eliminated Baierlein--Sharp--Wheeler (BSW) action of GR.  
These give rise to several branches of gravitational theories including GR on superspace and a theory of gravity on conformal superspace.         
This paper investigates the 3-space approach further, motivated both by the hierarchies of increasingly well-defined and weakened simplicity postulates present 
in all routes to relativity, and by the requirement that all the known fundamental matter fields be included.    

We further the study of configuration spaces of gravity-matter systems upon which BF\'{O}'s formulation leans.      
We note that in further developments the lapse-eliminated BSW actions used by BF\'{O} become impractical and require generalization.    
We circumvent many of these problems by the equivalent use of lapse-uneliminated actions, 
which furthermore permit us to interpret BF\'{O}'s formulation within Kucha\v{r}'s generally covariant hypersurface framework.   
This viewpoint provides alternative reasons to BF\'{O}'s as to why the inclusion of bosonic fields in the 3-space approach gives rise to minimally-coupled scalar fields, 
electromagnetism and Yang--Mills theory.  This viewpoint also permits us to quickly exhibit further GR-matter theories admitted by the 3-space formulation.    
In particular, we show that the spin-$\frac{1}{2}$ fermions of the theories of Dirac, Maxwell--Dirac and Yang--Mills--Dirac, all coupled to GR, are admitted by the generalized 
3-space formulation we present.  Thus all the known fundamental matter fields can be accommodated.  This corresponds to being able to pick actions for all these theories which 
have less kinematics than suggested by the generally covariant hypersurface framework.  For all these theories, Wheeler's thin sandwich conjecture may be \it posed\normalfont, 
rendering them timeless in Barbour's sense.

\end{abstract}

\maketitle

\section{Introduction}

Einstein \cite{Einstein}
`derived' his field equations (Efe's)\footnote{\noindent In this paper, 
spacetime tensors have lower-case Greek indices and 
space tensors have lower-case latin indices. 
Their barred counterparts are local Minkowski and Euclidean indices respectively.   
Bold capital Latin letters denote Yang--Mills internal indices. 
Capital Greek letters denote general indices. 
The indices $N$, $\bf N\normalfont$, $\xi$, $\chi$, $\Phi$ and $\Psi$, are reserved for other use.  
Round brackets surrounding more than one index of any type denote 
symmetrization
and square brackets denote antisymmetrization; indices which are not part of 
this (anti)symmetrization are
set between vertical lines. 
$g_{\alpha\beta}$ is the $(\epsilon + + +)$ spacetime metric, with 
determinant $g$,
where the signature $\epsilon = - s$ is -1 for (Lorentzian) GR, 0 for strong gravity and 
1 for Euclidean GR.  $\nabla_{\alpha}$ is the spacetime covariant derivative, 
$D_a$    is the   spatial covariant derivative and
$D^2$    is the spatial Laplacian.  
$\mbox{\bf R\normalfont}_{\alpha\beta}$ is the spacetime Ricci tensor,
$\mbox{\bf R\normalfont}$ is the spacetime Ricci scalar,
$\mbox{\bf G\normalfont}_{\alpha\beta}$ is the spacetime Einstein tensor and
$\mbox{\bf T\normalfont}_{\alpha\beta}$ is the energy-momentum tensor.
$h_{ab}$ is the metric on a spatial hypersurface, with determinant $h$. 
$p_{ab}$ is its conjugate momentum, with trace $p$.    
$R_{ab}$ is the spatial Ricci tensor and $R$ the spatial Ricci scalar.}
\be
\mbox{\bf G\normalfont}_{\alpha\beta}
= \mbox{\bf R\normalfont}_{ \alpha \beta } -
\frac{1}{2} g_{\alpha\beta }\mbox{\bf R\normalfont}
= 8 \pi\mbox{\bf T\normalfont}_{ \alpha \beta }^{\mbox{\scriptsize 
Matter\normalsize}}
\label{efes}
\ee
by demanding general covariance (GC) and the Newtonian limit; 
the conservation of energy-momentum requires $\nabla^{\alpha}\mbox{\bf{G}\normalfont}_{\alpha\beta} = 0$. 
Along with these physical considerations, 
Cartan \cite{Cartan} proved that the derivation requires the following mathematical simplicities: 
that $\mbox{\bf G\normalfont}_{\alpha\beta}$ contains at most second-order derivatives and is linear in these. 
The Efe's may also be obtained from the Einstein--Hilbert action \cite{Weyl}
\be
S_{\mbox{\scriptsize EH\normalsize}} = \int d^4x \sqrt{-g}(\mbox{\bf 
R\normalfont}
+ \mbox{\sffamily L\normalfont}_{\mbox{\scriptsize Matter\normalsize}});
\label{EinsteinHilbert}
\ee
an equivalent proof for actions was given by Weyl \cite{Weyl}.  Lovelock \cite{Lovelock} has shown that the linearity assumption is 
unnecessary in dimension $D \leq 4$. 

Arnowitt, Deser and Misner (ADM) \cite{ADM} split the spacetime metric as follows 
\be
\begin{array}{ll}
g_{\alpha\beta} = \left(\begin{array}{ll} \xi_k\xi^k - N^2   &    \xi_j    
\\
                                           \xi_i            &    h_{ij}
                   \end{array}\right),
&
g^{\alpha\beta} = \left(\begin{array}{ll} -\frac{1}{N^2}  & 
\frac{\xi^j}{N^2}  \\
                                          \frac{N^i}{N^2} & h^{ij} - 
\frac{\xi^i\xi^j}{N^2}
                    \end{array}\right)

\end{array}
\label{ADMsplit}
\ee
and rearranged the action ($\ref{EinsteinHilbert}$) into the Hamiltonian form 
\bea
S_{\mbox{\scriptsize ADM\normalsize}} = \int \mbox{d}t \int \mbox{d}^3x ( 
p^{ij}\dot{h}_{ij}- N{\cal H} - \xi^i{\cal H}_{i})
\label{ADM} \\
{\cal H} \equiv G_{ijkl}p^{ij}p^{kl} - \sqrt{h}R = 0 ,
\label{ham} \\
{\cal H}_i \equiv -2D_j{p_i}^j = 0 ,
\label{mom}
\eea
up to a divergence term. 
The lapse $N$ and shift $\xi_i$ have no conjugate momenta. 
Thus the true gravitational degrees of freedom in GR 
are contained in Riem, the space of Riemannian 3-metrics on a fixed topology 
taken here to be closed and without boundary. 
But the true degrees of freedom are furthermore subjected to the 
Hamiltonian and momentum constraints ${\cal H}$ and ${\cal H}_i$ respectively. 
If one can quotient out the 3-diffeomorphisms 
(which are generated by $\xi_i$), 
one is left with 
\be
\mbox{\scriptsize\{Superspace\}\normalsize} =
\frac {\mbox{\scriptsize\{Riem\}\normalsize}}
{\mbox {\scriptsize\{3-Diffeomorphisms\}\normalsize} },
\ee
which has naturally defined on it the DeWitt supermetric
$G_{ijkl} = \frac{1}{\sqrt{h}}\left(h_{i(k|}h_{j|l)} - 
\frac{1}{2}h_{ij}h_{kl}\right)$
present in the remaining constraint, ${\cal H}$.

Wheeler listed six routes to GR in 1973 \cite{MTW}. 
The first is Einstein's (plus simplicity postulate upgrades). 
The second is Hilbert's from (\ref{EinsteinHilbert}). 
The third and fourth are the two-way working 
between (\ref{EinsteinHilbert}) and (\ref{ADM}, \ref{ham}, \ref{mom}); these 
will concern us in this paper as the arena for Wheeler's question \cite{Battelle}: 
``If one did not know the Einstein--Hamilton--Jacobi equation,
how might one hope to derive it straight off from plausible first principles, 
without ever going through the formulation of the Einstein field equations 
themselves?".  The fifth and sixth routes mentioned are the Fierz--Pauli spin-2 field in an 
unobservable flat background \cite{FP} and Sakharov's idea that gravitation is 
the elasticity of space that arises from particle physics \cite{Sak}.   
One could add some more recent routes to Wheeler's list, such as from the 
closed string spectrum \cite{GSW}, and  the interconnection with Yang--Mills 
phase space in the Ashtekar variables approach \cite{Ashtekar}. 
Among these routes we distinguish three types: 
to relativity alone, to relativity with all known fundamental matter fields `added on', 
and to genuinely unified theories 
(whether partial such as already-unified Rainich--Misner--Wheeler theory 
\cite{Rainich}, Kaluza--Klein theory \cite{KK} and the Weyl  
gravitoelectromagnetic theory \cite{Weyltheory}, 
or total such as string theory). 
Finally, some routes will lead to modifications of GR, 
such as higher derivative theories or Brans--Dicke (BD) theory \cite{BD} 
(it is debateable whether string theory reproduces GR 
since string theory has a BD or `dilatonic' coupling). 
Simplicity postulates may be seen as a means of uniquely prescribing GR but 
there is no reason why nature should turn out to be simple in these ways. 
 
The original `seventh route to relativity' 
partial answer to Wheeler's question was given by Hojman, Kucha\v{r} and
Teitelboim (HKT) \cite{HKT}. As Wheeler suggested,  they attached importance
to an embeddability condition, which presupposes 4-dimensionally GC spacetime.
However, recently Barbour, Foster and \'{O} Murchadha
(BF\'{O}) \cite{BOF1} have provided a different partial answer \it without \normalfont
this presupposition.  In this paper, we study whether this
is (or can be made) satisfactory,  and how it compares to the HKT answer.

HKT required the `representation postulate':  that ${\cal H}$ and ${\cal H}_i$  be such that they close in the same way as the algebra of
deformations of a spatial hypersurface  embedded in a $( - + + + )$ Riemannian spacetime.  
This algebra is the Dirac Algebra, 
\be
\begin{array}{cc}
\{{\cal H}(x), {\cal H}(y)\} = {\cal H}^i(x)\delta_{,i}(x,y) + {\cal 
H}^i(y)\delta_{,i}(x,y)\\
\{{\cal H}_i(x), {\cal H}(y)\} = {\cal H}(x)\delta_{,i}(x,y)\\
\{{\cal H}_i(x), {\cal H}_j(y)\} = {\cal H}_i(y)\delta_{,j}(x,y) + {\cal 
H}_j(x)\delta_{,i}(x,y), 
\label{Algebra}
\end{array}
\ee
where $\{ \mbox{ } ,\}$ denote Poisson brackets.  
Their working is subject to the assumption that the evolution is path-independent, 
which means that the spacetime containing the hypersurface is foliation-invariant; 
this is the embeddability assumption. 
Their further time-reversal assumption is removed in \cite{Kuchar74}. 
Weakening their ans\"atze in stages (c.f the earlier \cite{Kucharearly}), 
they obtain as results that ${\cal H}$ must be ultralocal\fn{Ultralocal means 
no dependence on spatial derivatives.} and quadratic in its momenta 
and at most second order in its spatial derivatives (see however Sec II.F).  

The hope that pure geometrodynamics is by itself a total unified theory has largely been abandoned.  
So asking about ${\cal H} = 0$, which corresponds to the Einstein--Hamilton--Jacobi equation 
[by substituting $p^{ij} = \frac{\pa S}{\pa h_{ij}}$, for Jacobi's principal function $S$, in (\ref{ham})]
translates to asking about $^{\Psi}{\cal H} = 0$, including all the known fundamental matter fields, $\Psi$.  
We can now assess whether any first principles are truly plausible by seeing if they extend 
from a route to relativity alone to a route to relativity with all the known fundamental matter fields `added on'. 
The idea of the representation postulate extends additively (at least naively) 
to matter contributions to ${\cal H}$ and ${\cal H}_i$. 
Teitelboim \cite{Teitelboim} provided a partial extension of HKT's work to include 
electromagnetism, Yang--Mills theory and supergravity.  
One must note the absence of spin-$\frac{1}{2}$ fields from this list \cite{Wheeler}. 

In contrast, BF\'{O} require mere closure in place of closure as the Dirac Algebra. 
The strength of their method comes from the generalized Hamiltonian dynamics of Dirac \cite{Dirac}, 
which is taken further to provide a highly restrictive scheme based on exhaustion (see \cite{AB} for an account).  
They consider actions constructed according to two principles: best matching and local square roots (see below).  

The idea of BF\'{O}'s 3-space approach is to seek for laws of nature that have a relational form.  This is taken to mean that relative configurations alone are meaningful and 
that the time label is to play no role in the formulation.  The former is achieved by working indirectly with the relative configuration space via best matching.  The latter 
is emphasized by working with a manifestly reparametrization-invariant Jacobi-type square root action (see Sec II).  Furthermore it is chosen to have a local square root (see 
below).  Then the constraints of GR arise as direct consequences of the implementation of these two principles.  The 3-space approach advocates a space rather than spacetime 
ontology.  Rather than being presupposed, 4-dimensional general covariance and the spacetime form of the laws of nature is emergent in the 3-space approach.  We now carefully 
state the two principles for a class of actions from which GR will emerge as essentially singled out.  

\noindent
\bf 1 \normalfont :
the universal method of {\it{best matching}} \cite{BB, B94I, BOF1} is used
to implement the {\it{3-dimensional}} diffeomorphism invariance
by correcting the bare velocities of all bosonic fields $B$ according to the rule
$\dot{B} \longrightarrow \dot{B} - \pounds_{\xi}B$.\fn{$\lambda$ is the label along curves in superspace;
$\frac{\pa} {\pa \lambda}$ is denoted by a dot.  
$\pounds_{\xi}$ is the Lie derivative w.r.t $\xi_i$.}  For any two 3-metrics on 3-geometries $\Sigma_1$, $\Sigma_2$, this corresponds to keeping the coordinates of 
$\Sigma_1$ fixed whilst shuffling around those of $\Sigma_2$ until they are as `close' as possible to those of $\Sigma_1$.

\noindent
\bf 2 \normalfont: a {\it{local}} square root 
(taken at each space point before integration over 3-space) is used.  
Thus the pure gravity actions considered are of Baierlein--Sharp--Wheeler (BSW) 
\cite{BSW} type,
\be
S_{\mbox{\scriptsize BSW \normalsize}}
= \int \mbox{d}\lambda \int \mbox{d}^3x \sqrt{h} \sqrt{s R + \Lambda}
\sqrt{T_{\mbox{\scriptsize W\normalsize}}},
\label{ASBSW}
\ee
where $\Lambda$ is a cosmological constant.  

Writing this form amounts to applying a temporary simplicity postulate

\noindent
\bf 3 \normalfont:
the pure gravity action is constructed with
at most second-order derivatives\footnote{Furthermore, none of the
higher-order derivative potentials considered
by BF\'{O} turn out to be dynamically consistent (but see Sec II.F).} in the potential, 
and with a homogeneously quadratic best-matched kinetic term 
\be
T_{\mbox{\scriptsize W\normalsize}} = \frac{1}{\sqrt{h}}G_{\mbox{\scriptsize W\normalsize}}^{abcd}
(\dot{h}_{ab} - 2D_{(a}\xi_{b)}) ( \dot{h}_{cd} - 2D_{(c}\xi_{d)}),
\ee
where $G^{ijkl}_{\mbox{\scriptsize W\normalsize}} = \sqrt{h}(h^{ik}h^{jl} - Wh^{ij}h^{kl})$, 
$W \neq \frac{1}{3}$, 
is the inverse of the most general (invertible) ultralocal
supermetric \cite{DeWitt}, 
$G^{\mbox{\scriptsize X\normalsize}}_{abcd} = \frac{1}{\sqrt{h}}(h_{ac}h_{bd} - \frac{X}{2}h_{ab}h_{cd})$ 
for $X = \frac{2W}{3W - 1}$.

Setting $2N = \sqrt{T_{\mbox{\scriptsize W \normalsize}}/(sR + \Lambda)}$,
the gravitational momenta are 
\be
p^{ij} = \frac{\partial\mbox{\sffamily{L}\normalfont} }{ \partial\dot{h}_{ij}} =
\frac{\sqrt{h}}{2N}(h^{ic}h^{jd} - {W}h^{ij}h^{cd})(\dot{h}_{cd} - 2D_{(c}\xi_{d)}).
\ee
The primary constraint 
\be
{\cal H } \equiv -\sqrt{h}(sR + \Lambda) + \frac{1}{\sqrt{h}}
\left(
p^{ij}p_{ij} - \frac{X}{2}p^2 \right)  = 0
\label{GRHam}
\ee
then follows merely from the form of the Lagrangian.
In addition, variation of the action with respect to $\xi_i$
leads to a secondary constraint which is the usual momentum constraint 
(\ref{mom}).

The propagation of ${\cal H}$ gives \cite{SAnderson} 
$$
\dot{{\cal H}} = \frac{s}{N}D^i(N^2{{\cal H}_i})
+ \frac{(3X - 2)Np}{2\sqrt{h}}{\cal H}
+ \pounds_{\xi}{\cal H} 
$$
\be
\mbox{ } \mbox{ } \mbox{ } \mbox{ } \mbox{ } + \frac{2s(1 - X)}{N}D_a\left(N^2 D^ap\right).
\ee
We require this to vanish in order to have a consistent theory. 
The first 3 terms of this are said to \it vanish weakly \normalfont 
in the sense of Dirac \cite{Dirac}, i.e. they vanish by virtue of the constraints 
${\cal H}$, ${\cal H}_i$.  The last term has a chance to vanish in three ways, 
since it has three factors which might be zero.  
Constraints must be independent of $N$, so the third factor means that 
$p/\sqrt{h} = constant$.
We require this new constraint to propagate also, but this leads to the lapse being 
nontrivially fixed by a constant mean curvature (CMC) slicing equation.  
So, for $s \neq 0$, this forces us to have the DeWitt ($W = 1$)  
supermetric of relativity, which is BF\'{O}'s `Relativity Without Relativity' result.  

But there is also the $s = 0$ possibility regardless of which supermetric is 
chosen \cite{SAnderson},  which is a generalization of strong gravity 
\cite{SGlit}.  The HKT program would discard this since it is 
not a representation of the Dirac Algebra (although Teitelboim did study strong gravity \cite{SGlit}).
However, the strong gravity theories meet the 3-space approach's immediate 
criteria in being dynamically consistent theories of 3-geometries.  In this case the theories at most represent nature near singularities 
(although one can expand about them to obtain GR and Brans--Dicke theory) but it does illustrate that the 3-space approach is a fruitful 
constructive scheme for alternative theories.

Indeed, Barbour and \'{O} Murchadha (BO) found alternative conformal theories
\cite{conformal} which are being reformulated by
Anderson, Barbour, Foster and \'{O} Murchadha \cite{Foster}
using a new `free end-point' variational principle \cite{partI, Foster}.  \it Conformal gravity \normalfont has the action
$$
S_{\mbox{\scriptsize C\normalsize}} =
\int \mbox{d}\lambda
\frac{\int \mbox{d}^3x\sqrt{h}\phi^4
\sqrt{s
\left(
R - \frac{8D^2\phi}{\phi}
\right) +
\frac{   \Lambda\phi^4   }{   V(\phi)^{  \frac{2}{3}}   } }
\sqrt{T_{\mbox{\scriptsize C\normalsize}}}}
{  V(\phi)^{\frac{2}{3}}  },
$$
\be
\mbox{   volume } V = \int \mbox{d}^3x\sqrt{h}\phi^6 
\ee
$$
T_{\mbox{\scriptsize C\normalsize}} = 
\frac{1}{\sqrt{h}}G_{(\mbox{\scriptsize W \normalsize} = 0)}^{abcd}
(\dot{h}_{ab} - \pounds_{\xi}h_{ab} + \frac{ 4(\dot{\phi} - 
\pounds_{\xi}\phi)}{\phi}h_{ab})\times
$$
\be
(\dot{h}_{cd} - \pounds_{\xi}h_{cd} + \frac{4(\dot{\phi} - 
\pounds_{\xi}\phi)}{\phi}h_{cd}),
\label{BOaction}
\ee
which is consistent for $s=1$ because it circumvents the above
argument about the third factor by independently guaranteeing
a new slicing equation for the lapse.  Despite its lack of GC, 
conformal gravity is very similar to GR in the sense that the 
true configuration space of GR is 
\cite{Yorklit}
$$
\mbox{\scriptsize CS $+$ V\normalsize}
\equiv \{\mbox{\scriptsize Conformal Superspace $+$ Volume\normalsize}\} = 
$$
\be
\frac{\{\mbox{\scriptsize Riem\normalsize}\}}
{\{\mbox{\scriptsize 3-Diffeomorphisms\normalsize}\}
\{\mbox{\scriptsize Volume-preserving Weyl transformations\normalsize}\}}
\label{CSplusV}
\ee
and conformal gravity arises by considering instead
$$
\{\mbox{\scriptsize Conformal Superspace\normalsize}\}
= 
$$
\be
\frac{\{\mbox{\scriptsize Riem\normalsize}\}}
{\{\mbox{\scriptsize 3-Diffeomorphisms\normalsize}\}
\{\mbox{\scriptsize Weyl transformations\normalsize}\}}.
\label{CS}
\ee
This has an infinite number of `shape' degrees of freedom
whereas there is only one volume degree of freedom.
Yet removing this single degree of freedom
changes one's usual concept of cosmology,
and ought to change the problems associated with the quantization of the theory 
(by permitting the use of a positive-definite inner product 
and a new interpretation for ${\cal H}$) \cite{Foster}. 
Setting $s = 0$ in (\ref{BOaction}) gives strong conformal gravity.  
One arrives at a further \it CS+V \normalfont 3-space theory if one chooses to work on
(\ref{CSplusV}) instead of (\ref{CS}) \cite{Foster, ABFKO} whilst retaining a
fundamental slicing from the use of free-end-point variation.

To mathematically distinguish GR from these other theories, we use

\noindent
\bf 4 \normalfont: the theory is not conformally invariant, is obtained by 
conventional variation and has signature $\epsilon = -s = -1$.

\noindent The author's future strategy will involve seeking to overrule
these alternative theories by thought experiments and use of current
astronomical data, which would tighten the uniqueness of GR as a viable 
3-space theory \it on physical grounds\normalfont.
If such attempts persistently fail, these theories will become established as serious 
alternatives to GR.  So far the theories appear consistent with the GR solar system tests, 
and the CS+V theory will inherit the standard cosmology from GR.  

BF\'{O} furthermore considered `adding on' matter to the 3-geometries,\fn{We contest 
BF\'{O}'s speculation that the matter results might lead to unification in Sec IV.}
subject to the simplicity postulate 

\noindent
\bf 5 \normalfont: the matter potential has at most first-order derivatives and the kinetic term is  
ultralocal and homogeneous quadratic in the velocities.  Apart from the homogeneity, this parallels Teitelboim's matter assumptions \cite{Teitelboim}.    

\noindent 
One then discovers in the GR case that the lightcone is universal for bosons,
a single 1-form obeys Maxwell's electrodynamics, and sets of interacting
1-forms obey Yang--Mills theory \cite{AB}.
All these 1-forms have turned out to be massless.
Considering a 1-form and scalars simultaneously leads to $U(1)$ gauge theory \cite{BOF2}.
The GR matter results carry over to conformal gravity \cite{Foster}.

We sharpen the understanding of what the 3-space approach is because we are interested 
in why the impressive collection of results in the GR case above arises in BF\'{O}'s 
approach.  We seek for tacit simplicity postulates, survey which assumptions may 
be weakened and assess the thoroughness and plausibility of BF\'{O}'s principles, 
results and conjectures.  We thus arrive at a number of variations of the 3-space approach.
We stress that this is not just about improving the axiomatization.
We must be able to find a version that naturally accommodates spin-$\frac{1}{2}$ fermions
coupled 1) to GR if the 3-space approach is to provide a set of plausible first principles 
for GR 2) to conformal gravity if this is to be a viable alternative.  
Barbour's work \cite{B94I, Endoftime} has been critically discussed by Butterfield \cite{Butterfield} and by 
Smolin \cite{Smolin} largely from a philosophical point of view.  In contrast, this paper 
discusses (and extends) BF\'{O}'s continuation of this work from a more technical point of view.

In Sec II, we argue that the BSW principle \bf 2 \normalfont is problematic.
First, Barbour's use of it draws inspiration from the Jacobi formulation of mechanics, 
but in Sec II.A we point out that the Jacobi formulation itself has limitations 
and a significant generalization.  Furthermore in Secs II.B-D 
we point out that the differences between the BSW and Jacobi actions are important.  
Overall, this gives us the `conformal' problem in Sec II.C, and
the `notion of distance' problem in Sec II.D.
Second, should the notion of `BSW-type theories' not include all the theories 
that permit the BSW elimination process itself?  But when we perform this 
including fermions in Sec II.E, we find that we obtain not the BSW form but rather its generalization.    
Thus the inclusion of fermions will severely complicate the use of 
exhaustive proofs such as those in \cite{BOF1, AB}.   We furthermore 
point out that the usual higher derivative theories are not being excluded by 
BF\'{O}  in Sec II.F.  
These last two subsections include discussion of their HKT counterparts.

In Sec III, we formalize the second point above by showing that we could just as well use lapse-uneliminated actions for GR and conformal gravity.  
For GR, these actions may be studied within 
Kucha\v{r}'s GC hypersurface framework \cite{Kucharlit}.  
This framework brings attention to \it tilt \normalfont and 
\it derivative coupling \normalfont complications in general (Sec IV.A), 
which are however absent for the minimally-coupled scalar, 
and `accidentally absent' for the Maxwell and Yang--Mills 1-forms, 
which are what the 3-space approach picks out.  But tilt is present for the massive 
(Proca) analogues of these 1-forms. 
We deduce the relation between tilt and the existence of a generalized BSW form.  
In Sec IV.B we counter BF\'{O}'s hope that \it just \normalfont the known fundamental matter fields are 
being picked out by the 3-space approach, by showing that the massless 2-form is also compatible.  
In Sec IV.C, we find alternative reasons why the Maxwell 1-form is singled out by the 
3-space approach, from the point of view of the hypersurface framework.  
We end by explaining out the complications that would follow were one to 
permit derivative-coupled 1-forms.  

In Sec V.A, we point out that it is consistent to take the bosonic sector of nature  
to be far simpler than GC might have us believe:  best matching suffices for its  
construction. An alternative scheme to \bf 1 \normalfont using `bare' rather than best-matched velocities 
to start off with is discussed, in which ${\cal H}$ gives rise to all the other 
constraints as integrability conditions.  In Sec V.B, we show how all these results also hold true upon inclusion of spin-$\frac{1}{2}$ fermions.             
Sec V.C lists further research topics for fermions in the light of the advances made in this paper.  

\section{Problems with the Use of BSW Actions}

\subsection {Insights from Mechanics}

Suppose the Lagrangian\footnote{ Newtonian time is denoted by $t$ whilst $\tau$ is a parameter.  Dot is used for $\frac{\pa} {\pa t}$
in mechanics workings and dash for $\frac{\pa} {\pa \tau}$. $\hat{\Delta}$
takes $1$ to $n$ and $\Delta$ takes $1$ to $(n - 1)$; $n$ is not to be summed over.
$q_{\hat{\Delta}}$ are configuration variables with
conjugate momenta $p^{\hat{\Delta}}$.}
\be
\mbox{\sffamily L\normalfont}(q_{\hat{\Delta}}, \dot{q}_{\hat{\Delta}})
= 
\frac{1}{2}M^{\hat{\Delta}\hat{\Gamma}}(q_{\hat{\Pi}})\dot{q}_{\hat{\Delta}}\dot{q}_{\hat{\Gamma}}
- V( q_{\hat{\Pi}} )
\label{mechanicalL}
\ee
does not depend on $q_n$.  Then $q_n$ is a \it cyclic \normalfont variable and its Euler--Lagrange equation yields 
$p^n \equiv \frac {\pa\mbox{\sffamily\scriptsize L\normalsize\normalfont}} {\pa \dot{q}_n}  = c^n$, a constant.
Then the Lagrangian may be modified to $\bar{\mbox{\sffamily L\normalfont}}(q_{\Delta}, \dot{q}_{\Delta}) \equiv \mbox{\sffamily L\normalfont} - c^n\dot{q}_n$
using the equation for $p^n$ to eliminate $\dot{q}_n$; 
this is known as \it Routhian reduction \normalfont.

Next, observe that $q_n$ may be taken to be the time $t$
in a conservative mechanical system; we regard the
$q_{\Delta}$ and $t$ as functions of the parameter $\tau$.
Then the action takes the parametrized form
\be
S = \int_{\tau_1}^{\tau_2}
\mbox{\sffamily L\normalfont}(q_{\Delta}, 
\frac{q^{\prime}_{\Delta}}{t^{\prime}})t^{\prime}\mbox{d}\tau,
\label{unelparam}
\ee
and the equation for $p^t$ may be used to eliminate $t^{\prime}$
from this by Routhian reduction.  One thus obtains the Jacobi 
action
\be
S_{\mbox{\scriptsize J\normalsize}} = \int_{\tau_1}^{\tau_2}\sqrt{2(E - 
V)}\mbox{d}\sigma,
\ee
where $E \equiv c^t$ is the total energy and $\mbox{d}\sigma^2$ is the line element associated 
with the
Riemannian metric $M_{\Gamma\Delta}$ of the configuration
space $Q$ of the configuration variables $q_{\Delta}$.
Minimization of this integral is \it Jacobi's principle \normalfont 
\cite{Lanczos}.
There is then a conformally-related line element
\be
\mbox{d}\tilde{\sigma}^{2} = (E - V)\mbox{d}\sigma^2
\label{crel}
\ee
with respect to which the motions of the system are geodesics.
The point of this method is the reduction of mechanics problems to the study
of well-known geometry.

However, the Jacobi principle in mechanics has a catch: the conformal
factor is not allowed to have zeros.
If it does then the conformal transformation is only valid in regions where
there are no such zeros.
These zeros are physical barriers in mechanics.
For they correspond to zero kinetic energy by the conservation of energy 
equation.  As the configuration space metric is positive-definite, this means that the 
velocities must be zero there, so the zeros cannot be traversed.

The Lagrangian (\ref{mechanicalL}) is restricted to have a kinetic term
homogeneously quadratic in the velocities.
Let $\mbox{\sffamily L\normalfont}(q_{\hat{\Delta}}, \dot{q}_{\hat{\Delta}})$
be instead a completely general function.  Then 
\be 
S = \int_{\tau_1}^{\tau_2}\mbox{\sffamily L\normalfont}(q_{\Delta}, 
\frac{q^{\prime}_{\Delta}}{t^{\prime}})t^{\prime}\mbox{d}\tau \equiv  
\int_{\tau_1}^{\tau_2}{\cal L}(q_{\hat{\Delta}}, q^{\prime}_{\hat{\Delta}})d\tau 
\ee
may be modified to 
\be
S_{\mbox{\scriptsize J\normalsize}} 
= \int_{\tau_1}^{\tau_2}\bar{\cal L}(q_{\Delta}, q^{\prime}_{{\Delta}}) \mbox{d}\tau
\label{Jacgen}
\ee
by Routhian reduction, where $\bar{\cal L} = F$, some homogeneous linear function of the $q^{\prime}_{\Delta}$ \cite{Lanczos}.    
For example, $F$ could be a \it Finslerian metric function \normalfont from 
which we could obtain a Finslerian metric $f_{\Gamma\Delta} = \frac{1}{2}
\frac{  \pa^2  }{  \pa q^{\prime}_{\Gamma} \pa q^{\prime}_{\Delta}  }F^2$,
provided that $F$ obeys further conditions \cite{finsler} including the nondegeneracy of $f_{\Gamma\Delta}$.
So in general the `geometrization problem' of reducing the motion of a
mechanical system to a problem of finding geodesics involves more than
the study of Riemannian geometry.

To some extent, there is conventional freedom in the choice of
configuration space geometry, since we notice that standard manoeuvres
can alter whether it is Riemannian.  This is because one is free in how many
redundant configuration variables to include, and in the character of those
variables (for example whether they all obey second-order Euler--Lagrange 
equations).

As a first example, consider the outcome of the Routhian reduction of (\ref{mechanicalL}) more carefully:
$$
\bar{\mbox{\sffamily L\normalfont}}(q_{\Delta}, \dot{q}_{\Delta})
= \frac{1}{2}\left( M^{\Gamma\Delta} - \frac{M^{\Delta n}M^{\Gamma 
n}}{M^{nn}}\right)
\dot{q}_{\Delta}\dot{q}_{\Gamma} 
$$
\be
+ \frac{c^nM^{\Delta n}}{M^{nn}}\dot{q}_{\Delta} - 
\bar{V},
\label{careRouth}
\ee
where $\bar{V}$ is a modified potential.
So Routhian reduction can lead to non-Riemannian geometry,
on account of the penultimate `gyroscopic term'
\cite{Lanczos}, which is linear in the velocities.
We consider the reverse of this procedure as a possible means of arriving at 
Riemannian geometry to describe systems with linear and quadratic terms.
We observe that if the linear coefficients depend on configuration 
variables,
then in general the quadratic structure becomes contaminated with these 
variables.

As a second example, higher-than-quadratic systems may be put into quadratic 
form
by \it Ostrogradsky reduction \normalfont \cite{ostrograd}, at the price of introducing extra
configuration variables.

We finally note the ordering of the summation and the square root in
\be
\mbox{d}\sigma = \sqrt{\sum_{\Delta, \Gamma = 1}^{n - 
1}
\tilde{M}^{\Gamma\Delta}\dot{q}_{\Gamma} \dot{q}_{\Delta} },
\ee
which we refer to as the `good' or `global square root' ordering.

\subsection{The BSW Formulation of GR}

GR is an already-parametrized theory.  This is because
the ADM action (\ref{ADM}) (generalized to arbitrary $s$ and $\Lambda$ at no extra cost) may be rewritten in the Lagrangian form
$$
S = \int \mbox{d}\lambda\int \mbox{d}^3x \sqrt{h}N\mbox{\sffamily L\normalfont}
(h_{ab}, \dot{h}_{ab}; \xi_i; N)
$$
\be
= \int \mbox{d}\lambda \int \mbox{d}^3x\sqrt{h}N\left(\Lambda + sR
+ \frac{T_{\mbox{\scriptsize g\normalsize}}(\kappa_{ij})}{4N^2}\right),
\label{BSWmethod}
\ee
[c.f (\ref{unelparam})] where
\bea
T_{\mbox{\scriptsize g\normalsize}} = \kappa_{ij} \kappa^{ij} - \kappa^2, &
\kappa_{ij} = \dot{h}_{ij} - 2D_{(i}\xi_{j)}.
\eea
Then (specifically following BSW \cite{BSW} or in analogy with Jacobi)
extremization w.r.t. $N$ gives $N = \pm \sqrt{T_{\mbox{\scriptsize g\normalsize}}/(\Lambda + sR)}$,
which may be used to \it algebraically \normalfont eliminate $N$ from
(\ref{BSWmethod}).  Thus one arrives at the BSW action
\be
S_{\mbox{\scriptsize BSW\normalsize}} = \int \mbox{d}\lambda \int \mbox{d}^3x \sqrt{h} 
\sqrt{(\Lambda +sR)T_{\mbox{\scriptsize g\normalsize}}}.
\label{Bashwe}
\ee

Although this looks similar to the Jacobi action in mechanics,
there are important differences.
First, the GR configuration space is infinite-dimensional;
with redundancies, one can consider it to be superspace.
The DeWitt supermetric is defined on superspace \it pointwise \normalfont.
By use of a 2-index to 1-index map $G_{abcd} \longrightarrow G_{AB}$, 
DeWitt represented his supermetric as a
$6\times 6$ matrix, which is $( - + + + + + )$ and thus indefinite 
\cite{DeWitt}.
As a special case, minisuperspace \cite{Misner} is the truncation of 
superspace
obtained by considering homogeneous metrics alone.
`Minisupermetrics' are $( - + + )$, thus they too are indefinite.
Second, the BSW action has the `bad' or `local square root' ordering.
Below, we first consider minisuperspace, for which this extra complication 
does not
arise, since by homogeneity the `good' Jacobi and `bad' BSW orderings are 
equivalent.

Finally, BSW's work led to the thin sandwich conjecture \cite{TSL, Giusan}, the solubility of which features as a caveat in BF\'{O}'s original paper.  
Being able to pose this conjecture for a theory amounts to being able to algebraically eliminate the lapse $N$ from its Lagrangian.  This implies that the theory is timeless 
in Barbour's sense \cite{B94I, Endoftime}.  The extension of the conjecture to include fundamental matter fields has only recently begun \cite{Giusan}.  This and other 
investigations are required to assess the robustness of the conjecture to different theoretical settings, to see if in any circumstances it becomes advantageous to base 
numerical relativity calculations on the algorithm which the conjecture provides.   

\subsection{Lack of Validity of the BSW Form}

In perfect analogy with mechanics (\ref{crel}),
there is a conformally-related line element,
$\mbox{d}\tilde{\sigma}^{2} = (\Lambda + sR)\mbox{d}\sigma^2$
in vacuo, for which the motion associated with (\ref{Bashwe}) is geodesic \cite{DeWitt70}.
But the observation in mechanics that such conformal transformations are 
only valid
in regions where the conformal factor is nonzero\fn{In GR,
these are regions for which $\Lambda + R < 0$ or for which $\Lambda + R > 
0$.
We also note that the sign of $\Lambda + R$ plays an important role in the
thin sandwich conjecture.} still holds for GR.
It is true that the details are different,
due to the indefiniteness of the GR supermetric.
This causes the zeros to be spurious rather than physical barriers 
\cite{Misner}.
For whilst a zero $z$ of the potential corresponds to a zero of the kinetic 
term
by virtue of the Hamiltonian constraint,
this now means that the velocity need be null, not necessarily zero, because of the indefiniteness.
Thus the motion may continue through $z$ `on the superspace lightcone',
which is made up of perfectly reasonable Kasner universes,
rather than grind to a halt.
Nevertheless, the conformal transformation used to
obtain geodesic motion is not valid, so it is questionable whether the BSW form 
is a `geodesic principle', if in general it describes conformally untransformed 
\it non-geodesic curves \normalfont for practical purposes.

To illustrate that the presence of zeros in the potential term is an
important occurrence in GR, we note that the Bianchi IX solution has an infinity 
of such zeros as one approaches the cosmological singularity.
This is important because it is conjectured by Belinskii, Khalatnikov and Lifshitz 
(BKL) \cite{BKL} that the behaviour of Bianchi IX near the cosmological singularity is 
the generic behaviour of a cosmological solution to GR.
This sort of conjecture is acquiring numerical support \cite{Bergerlit}.
The above argument was originally put forward by Burd and Tavakol \cite{TB}
to argue against the validity of the use of the `Jacobi principle'
to characterize chaos in GR \cite{Sydlowski}.
Our point is that this argument holds against \it any \normalfont use, BF\'{O}'s 
included, of the BSW form in minisuperspace models of the early universe in GR.  

The way out of this argument that we suggest is to abstain from the 
self-infliction
of spurious zeros by not performing the conformal transformation in the first 
place,
thus abandoning the interpretation of the BSW form as a geodesic principle 
in GR.
Conformal gravity however is distinct from GR and has no cosmological 
singularity, 
so arguments based on the BKL conjecture are not applicable there.
Conformal gravity's zeros are real as in mechanics,
because  $T^{\mbox{\scriptsize C\normalsize}}$ is positive-definite, and
Barbour and \'{O} Murchadha use this to argue that topologies with $R < 0$ at any point are not allowed \cite{conformal}.

\subsection{The BSW Form is an Unknown Notion of Distance}

BF\'{O} called the local square root ordering `bad' because it gives one constraint 
per space point, which would usually render a theory trivial by overconstraining 
due to the ensuing cascade of secondary constraints.  
Yet GR contrives to survive this because of its hidden foliation invariance
\cite{BOF1}.  However Giulini \cite{Giusan} has pointed out another reason why the local
square root ordering is bad: it does not give rise to known geometry.
Below, we extend his finite-dimensional counterexample to the geometry being Finslerian.

The BSW form as a notion of distance provides as the `full metric' on superspace 
$$
\frac{1}{2}\frac{\pa^2}{\pa v^A(u)\pa v^B(w)}
S_{\mbox{\scriptsize BSW\normalsize}}^2 =
\left[
\tilde{G}_{AC}(u)\tilde{G}_{BD}(w) \right.
$$
\be
\left.
+ 2\delta^{(3)}(u, w)
\left(\frac{S_{\mbox{\scriptsize 
BSW\normalsize}}}{\sqrt{\tilde{G}_{BD}v^Bv^D }}
\tilde{G}_{A[B}\tilde{G}_{C]D}\right)(u)
\right]\hat{v}^A\hat{v}^C,
\ee
where $v^A \equiv \dot{h}^A = \dot{h}^{ab}$ by DeWitt's 2-index to 1-index 
map
and where hats denote unit `vectors'.
So in general, $\tilde{G}_{A[B}\tilde{G}_{C]D} = 0$ is a sufficient
condition for the full metric to be degenerate and hence not Finsler
(Giulini's example had a 1-dimensional $v^A$ so this always occurred).
But if $\tilde{G}_{A[B}\tilde{G}_{C]D} \neq 0$, the full metric is not a 
function
(both in the distributional and functional senses).
So using the BSW form as a notion of distance leads to unknown geometry,
so there is no scope for the practical application
of the BSW form as a geodesic principle.

This is to be contrasted with the global square root, for which the
above procedure gives instead (semi-)Riemannian geometry.  For minisuperspace, 
the local square root working presented does indeed collapse
to coincide with this global square root working, and the resulting (semi)Riemannian 
geometry is of considerable use in minisuperspace quantum cosmology \cite{Misner}.

There is also the issue DeWitt raised \cite{DeWitt70} 
that in the study of superspace one is in fact considering not single geodesics, 
but \it sheaves \normalfont of them.  This corresponds to all the different foliations of 
spacetime in GR, which leads to the problem of time in quantum gravity \cite{POT}.  
Thus there are two difficulties with applying BF\'{O}'s formulation of GR.
The first will still plague conformal gravity whereas the second is absent because there 
is a preferred lapse rather than foliation invariance.

\subsection {The Fermionic Contribution to the Action is Linear}

Since the kinetic terms of the bosons of nature
are also quadratic in their velocities, we can use the modifications
\bea
T_{\mbox{\scriptsize g\normalsize}}\longrightarrow T_{\mbox{\scriptsize 
g\normalsize}}
+ T_{\mbox{\scriptsize B\normalsize}},
& \Lambda +  sR \longrightarrow \Lambda + sR + U_{\mbox{\scriptsize 
B\normalsize}}
\eea
to accommodate bosonic fields $B_{\Delta}$ in a BSW-type action,
\be
S_{\mbox{\scriptsize B\normalsize}} = \int \mbox{d}\lambda
\int \mbox{d}^3x \sqrt{h} \sqrt{\Lambda + s R + U_{\mbox{\scriptsize 
B\normalsize}}}
\sqrt{T_{\mbox{\scriptsize g\normalsize}} + T_{\mbox{\scriptsize 
B\normalsize}}}
\label{BSWboson}.
\ee
This local square root encodes the correct Hamiltonian constraint for the
gravity--boson system.
Although the pointwise Riemannian kinetic metric is larger than
the DeWitt supermetric, in the case of minimally-coupled matter
it contains the DeWitt supermetric as an isolated block:
\be
\left(
\begin{array}{ll}
G_{AB}(h_{ab}) & 0 \\
0 & H^{\mbox{\scriptsize Matter\normalsize}}_{\Lambda\Sigma}(h_{ab})
\end{array}
\right)
\label{min}
\ee
If this is the case, it makes sense to study the pure gravity part by 
itself, which
is a prominent feature of almost all the examples studied in the 3-space approach.
We identify this as a tacit simplicity requirement,
for without it the matter degrees of freedom interfere with the 
gravitational ones,
so it makes no sense then to study gravity first and then `add on' matter
In Brans--Dicke (BD) theory, this is not immediately the case: this is an example 
in which
there are gravity-boson kinetic cross-terms $C_{A\chi}$ in the pointwise
Riemannian kinetic metric:
\be
\left(
\begin{array}{ll}
G^{\mbox{\scriptsize X\normalfont}}_{AB}(h_{ab}) & C_{A\chi}(h_{ab}, \chi) \\  C_{\chi B}(h_{ab}, \chi) &
H^{\mbox{\scriptsize Matter\normalsize}}_{\chi\chi}(h_{ab}, \chi)
\end{array}
\right)
\ee
where $\chi$ is the BD field and $X$ is related to the usual BD parameter $\omega$ by $\frac{X}{2} = \frac{\omega + 1}{2\omega + 3}$.
Thus, the metric and dilatonic fields form
\it together \normalfont a theory of gravity with 3 degrees of freedom.
However, this is a mild example of nonminimal coupling because redefinition 
of the metric and scalar degrees of freedom permits blockwise isolation of the form (\ref{min}).
More disturbing examples are considered below and in Sec IV.C.

We now begin to consider whether and how the 3-space formulation can 
accommodate
spin-$\frac{1}{2}$ fermionic fields, $F_{\Delta}$.  Following the strategy
employed above for bosons, the BSW working becomes
$$
S_{\mbox{\scriptsize F\normalsize}} 
= \int \mbox{d}\lambda\int \mbox{d}^3x \sqrt{h}NL(h_{ab}, \dot{h}_{ab}; \xi_i; N; 
F_{\Delta}, \dot{F}_{\Delta}) 
$$
$$
= \int \mbox{d}\lambda \int \mbox{d}^3x\sqrt{h}
\left[
N
\left(
\Lambda + s R + U_{\mbox{\scriptsize F\normalsize}}
+ \frac{T_{\mbox{\scriptsize g\normalsize}}(\kappa_{ij})}{4N^2}
\right)
\right.
$$
\be
\left. 
+ T_{\mbox{\scriptsize F\normalsize}}(\dot{F}_{\Delta})\right]
\label{SFtrue}
\ee
because $T_{\mbox{\scriptsize F\normalsize}}$ is linear in 
$\dot{F}_{\Delta}$.\fn{We see in Sec IV that the algebraic dependence on $N$ emergent from such decompositions requires rigorous justification.  
We provide this for (\ref{SFtrue}) in Sec V.B.}
Then the usual trick for eliminating $N$ does not touch 
$T_{\mbox{\scriptsize F\normalsize}}$,
which is left outside the square root:
\be
S_{\mbox{\scriptsize F\normalsize}} = \int \mbox{d}\lambda
\int \mbox{d}^3x \sqrt{h}\left(\sqrt{\Lambda + sR + U_{\mbox{\scriptsize
F\normalsize}}} \sqrt{T_{\mbox{\scriptsize g\normalsize}}} +
T_{\mbox{\scriptsize F\normalsize}}\right). \label{actualfermi}
\ee
The local square root constraint encodes
the correct gravity-fermion Hamiltonian constraint
\be
^{\mbox{\scriptsize F\normalsize}}{\cal H}
\equiv -\sqrt{h}(\Lambda + sR + U_{\mbox{\scriptsize F \normalsize}})
+ \frac{1}{\sqrt{h}}\left(p_{ij}p^{ij} - \frac{X}{2}p^2\right) = 0.
\ee
We postpone the issue of best matching (which is intertwined with gravity-fermion momentum constraint) until Sec. V.B.
Our concern in this section is the complication of the configuration space geometry due to the inclusion of fermions.

For now the elimination procedure is analogous not to the
Jacobi working but rather to its generalization (\ref{Jacgen}).
So even the pointwise geometry of the  gravity-fermion configuration space
is now compromised:  $\sqrt{\Lambda + s R + U_{\mbox{\scriptsize F\normalsize}}}
\sqrt{T_{\mbox{\scriptsize g\normalsize}}} + T_{\mbox{\scriptsize F\normalsize}}$
could sometimes be a Finslerian metric function.  
By allowing (\ref{SFtrue}), we are opening the door to all sorts of
complicated possible actions, such as

\noindent
1) $^k\sqrt{ G^{\Sigma_1 ... \Sigma_{k}}\dot{q}_{\Sigma_1} ... 
\dot{q}_{\Sigma_{k}} }$.

\noindent
2) Arbitrarily complicated compositions of such roots, powers and sums.

\noindent
3)More generally, $K_{\Delta}\dot{q}^{\Delta}$,
where $K_{\Delta}$ is allowed to be an
arbitrary function of not only the $q_{\Delta}$
but also of the $\Delta - 1$ independent ratios of the velocities.

\noindent
4) The above examples could all be Finslerian or fail to be so
by being degenerate.    
They could also fail to be Finslerian if the
$K_{\Delta}$ are permitted to be \it functionals \normalfont
of overall degree 0 in the velocities,
which we can take to be a growth of the local-global square root ambiguity.

We would therefore need to modify the BSW principle \bf 2 \normalfont to
a general BSW principle \bf 2G \normalfont that includes spin-$\frac{1}{2}$ fermions.
This amounts to dropping the requirement of the matter field kinetic term being 
homogeneously quadratic in its velocities, thus bringing \bf 5 \normalfont into alignment with Teitelboim's assumptions.  
We note that with increasing generality the possibility
of uniqueness proofs becomes more remote.
Although some aims of the 3-space approach
such as a full derivation of the universal light-cone would require some level of 
uniqueness proofs for spin-$\frac{1}{2}$ fermions, the author's strategy is to show in this paper that spin-$\frac{1}{2}$ fermions coupled to GR do possess a 3-space formulation 
and also to point out that the uniqueness results may have to be generalized in view of the generalization of the BSW form required in this section.   

Could we not choose to geometrize the gravity-fermion system as
a Riemannian geometry instead, by use of the reverse of Routhian reduction?
But the coefficients of the linear fermionic velocities in the
Einstein--Dirac system contain fermionic variables, so the resulting
Riemannian geometry's coefficients would contain the fermionic variables in addition 
to the metric.  We call such an occurrence a \it breach of the DeWitt structure\normalfont, since it means that contact is lost with DeWitt's study of the
configuration space of pure GR \cite{DeWitt, DeWitt70}.
So this choice also looks highly undesirable.

For 40 years the natural accommodation of spin-$\frac{1}{2}$ fermions in geometrodynamics \cite{Wheeler} 
has been a source of problems.   So this is a big demand on the 3-space approach, and one which must be met 
if the 3-space approach is truly to describe nature.  Our demands here are less than Wheeler's in \cite{Wheeler}: 
we are after a route to relativity with all matter `added on' rather than a complete unified theory.  
The HKT route appears also to be incomplete at this stage:  Teitelboim was unable to find a hypersurface deformation 
explanation for spin-$\frac{1}{2}$ fermions \cite{Teitelboim}.  
Thus when we began this work, all forms of the seventh route to relativity were 
incomplete w.r.t the inclusion of spin-$\frac{1}{2}$ fermions.  
In Sec V.B, we will point out the natural existence of GR--spin-$\frac{1}{2}$ theory 
within the 3-space approach.  

\subsection{Higher Derivative Theories}

We now argue against the significance of the preclusion of higher derivative theories 
by BF\'{O}.  For the precluded theories are easily seen \it not \normalfont to be 
the usual higher derivative theories.  There are two simple ways of noticing this.  
First, the primary constraints encoded by the BF\'{O} theories
with arbitrary $P(h_{ij}, h_{ij,k}, ...)$ will always be of the form
\be
\sqrt{h}{\cal H} = -\sqrt{h}P +\frac{1}{\sqrt{h}}\left(p_{ij}p^{ij} -
\frac{X}{2}p^2\right) = 0,
\ee
which is not what one gets for the usual higher derivative theories.
Second, BF\'{O}'s theories have fourth-order terms in their potentials but
their kinetic terms remain quadratic in the velocities, whilst the usual higher 
derivative theories' kinetic terms are quartic in the velocities.  
We argue that the mismatch of derivatives between $T$ and $P$ for
$P \neq sR + \Lambda$, overrules the theories from within the GC framework,
so BF\'{O} are doing nothing more than GC can do in this case.

It is not clear whether the usual higher derivative theories could be 
written in some generalized BSW form.
The form would either be considerably more complicated than that of pure GR or 
not exist at all.  Which of these is actually true should be checked case by case.  
We consider this to be a worthy problem in its own right by the final comment in Sec II.B, since  
this problem may be phrased as `for which higher derivative theories can the thin sandwich formulation be posed?'  
To illustrate why there is the possibility of nonexistence,
consider the simplest example, $\mbox{\bf R\normalfont} +
\alpha \mbox{\bf R\normalfont}^{2}$ theory.
The full doubly-contracted Gauss equation is
\be
\mbox{\bf R\normalfont}
= R - s(K_{ab}K^{ab} - K^2) + 2s D_a(n^b D_bn^a - n^a D_bn^b)
\ee
and, whereas one may discard the
divergence term in the 3 + 1 split of $\mbox{\bf R\normalfont}$, in the 3 + 
1 split
of $\mbox{\bf R\normalfont}^{2}$, this divergence is multiplied by 
$\mbox{\bf R\normalfont}$ and so cannot similarly be discarded.
So it is unlikely that the elimination of $N$ will be \it algebraic \normalfont
in such theories, which is a requirement for the BSW procedure.\fn{On the other hand, if $N$ occurs only linearly in the action then the variational equation for $N$ contains 
no $N$ and so cannot be used to eliminate $N$.  If $N$ occurs homogeneously in the action, then the variational equation for $N$ contains $N$ only as an overall factor and 
so cannot be used to eliminate $N$ either.  Also, it \it is \normalfont permissible for derivatives of $N$ to be present, so long as these terms belong to a total divergence 
which may then be discarded to leave an action depending only algebraically on $N$.  This might conceivably happen for some cases of higher derivative theories.  Finally note 
that the form in which $N$ appears in the action may change under changes of variables.}
Were this algebraic elimination possible, we would get more complicated 
expressions than the local square root form from it.
Indeed, higher derivative theories are known to have considerably more
complicated canonical formulations than GR \cite{hdl};
it is standard to treat them by a variant of Ostrogradsky reduction
adapted to constrained systems \cite{hdl}.

It is worth commenting that HKT's derivation of
${\cal H}$ being quadratic in its momenta
and containing at most second derivatives may also be interpreted as tainted,
since it comes about by restricting the gravity to have two degrees of freedom, as opposed to e.g. the three of 
$\mbox{\bf R\normalfont} + \alpha \mbox{\bf R\normalfont}^{2}$ theory or of Brans--Dicke theory.
Thus we do not foresee that any variant of the seventh route to relativity 
will be able to find a way round the second-order derivative assumption of the other routes.

\section{Lapse-uneliminated Variations on the 3-Space Approach}

We have seen that the interpretation of the BSW form
as a geodesic principle is subject to considerable complications,
and that it may obscure which theories are permitted or forbidden
in the 3-space approach.  We will now show that the use of the BSW form,
and consequently the problems with its interpretation,
may be circumvented by the use of lapse-uneliminated actions
because the content of GR is not affected by lapse elimination
(just as the Jacobi and Euler--Lagrange interpretations of mechanics are 
equivalent).
It is easy to show that the equations of motion that follow from
the $N$-uneliminated 3 + 1 `ADM' Lagrangian (\ref{BSWmethod})
are weakly equivalent to the BSW ones:
$$
\left(\frac{\partial p^{ij}} {\partial \lambda}\right)_{\mbox{\scriptsize 
ADM\normalsize}}
= \sqrt{h}N\left(h^{ij}\frac{sR + \Lambda}{2} - sR^{ij}\right) 
$$
$$
- \frac{2N}{\sqrt{h}}\left(p^{im}{p_m}^j -\frac{X}{2}p^{ij}p\right)
$$
$$
+ \frac{N}{2\sqrt{h}}h^{ij}\left(p_{ab}p^{ab} - \frac{X}{2}p^2 \right)
$$
$$
+ s\sqrt{h}(D^iD^jN -h^{ij} D^2 N) + \pounds_{\xi}p^{ij}
$$
$$
= \sqrt{h}N[h^{ij}(sR\mbox{+}\Lambda)\mbox{--}sR^{ij}] 
$$
$$
- \frac{2N}{\sqrt{h}}(p^{im}{p_m}^j -\frac{X}{2}p^{ij}p)
$$
$$
+ s\sqrt{h}( D^iD^jN - h^{ij}D^2 N)
+ \pounds_{\xi}p^{ij}
$$
$$
- \frac{N}{2}h^{ij}\left[\sqrt{h}(sR + \Lambda)
- \frac{1}{\sqrt{h}}\left(p_{ab}p^{ab} - \frac{X}{2}p^2 \right)\right]
$$
\be
= \left(\frac{\partial p^{ij}}{\partial \lambda}\right)_{\mbox{\scriptsize 
BSW \normalsize}}
+ \frac{N}{2}h^{ij}{\cal H},
\label{BSWtoADM}
\ee
and similarly when matter terms are included.  We use arbitrary $s$ and $W$ 
above
to simultaneously treat the GR and strong gravity cases.
The ADM propagation of the Hamiltonian constraint is slightly simpler than 
the BSW one,
\be
\dot{{\cal H}} = \frac{s}{N}D^i(N^2{\cal H}_i) + \pounds_{\xi}{\cal H}
\ee
for $W = 1$ or $s = 0$,
where it is understood that the evolution is carried out by the
ADM Euler--Lagrange equations or their strong gravity analogues.

We now check that using uneliminated actions does not damage the conformal 
branch of the
3-space approach.  The conformal gravity action (\ref{BOaction}) is 
equivalent to
\be
S = \int \mbox{d}\lambda
\frac
{\int \mbox{d}^3x\sqrt{h}N\phi^4
\left[
s\left(
R - \frac{ 8 D^2 \phi }{ \phi }
\right)
+ \frac{\Lambda\phi^4}{V^{\frac{2}{3}}(\phi)}
+ \frac{T_{\mbox{\scriptsize C\normalsize}}}{4N^2}
\right]}
{V(\phi)^{\frac{2}{3}}}
\ee
where the lapse is
$N = \frac{1}{2}
\sqrt{    \frac{   T_{\mbox{\scriptsize C\normalsize}}   }
{   s
\left(
R - \frac{ 8 D^2 \phi }{ \phi }
\right)
+ \frac{  \Lambda\phi^4  }{  V^{\frac{2}{3}}  }   }    }$.
The following equivalent of (\ref{BSWtoADM}) holds:
$$
\left(\frac{\partial p^{ij}}
{\partial\lambda}\right)_{\mbox{\scriptsize N-uneliminated\normalsize}} =
\left(\frac{\partial p^{ij}}
{\partial\lambda}\right)_{\mbox{\scriptsize N-eliminated\normalsize}}
$$
\be
+ h^{ij}
\left(
\frac{N{\cal H}^{\mbox{\scriptsize C\normalsize}}}{2}
- \frac{\sqrt{h}\phi^6}{3V}\int \mbox{d}^3x {   N {\cal H}^{\mbox{\scriptsize
C\normalsize}}   } \right)
\label{CBSWtoCADM}
\ee
for
\be
{\cal H}^{\mbox{\scriptsize C\normalsize}} \equiv
- \frac{  \sqrt{h}\phi^4   }{   V^{  \frac{2}{3}  }   }
\left[
s
\left( R - \frac{8D^2\phi}{\phi}
\right)
+ \frac{\Lambda\phi^4}{V^{\frac{2}{3}}}
\right]
+ \frac{V^{\frac{2}{3}}}{\sqrt{h}\phi^4}p^{ab}p_{ab}
\ee
the conformal gravity equivalent of the Hamiltonian constraint.

We now develop a strategy involving the study of lapse-uneliminated actions.
This represents a first step in disentangling Barbour's no time \cite{B94I, Endoftime}
and no scale \cite{partI, Foster} ideas.
It also permits us to investigate which standard theories exist
according to the other 3-space approach rules, by inspection of formalisms of these 
theories.  We could then choose to algebraically eliminate the lapse 
where
possible to show which of these theories can be formulated in the original 
BF\'{O}
3-space approach.  We emphasize that existence is by no means guaranteed:
some perfectly good GC formulations of theories are not best-matched, 
or do not permit a BSW reformulation because they can not be made to depend algebraically on the lapse.
Thus the uneliminated form can be used to help test whether the 3-space
approach is or can be made to be a satisfactory scheme for all of nature.

We can furthermore use this lapse-uneliminated formulation to interpret 
the GR branch of the 3-space approach within Kucha\v{r}'s hypersurface framework,
which has striking interpretational consequences, to which we now turn.

\section{The 3-Space Approach and the Hypersurface Framework}

\subsection{Nonderivatively Coupled 1-Forms}

In his series of four papers,
Kucha\v{r} \cite{Kucharlit} considers (I) the deformation of a hypersurface,
(II) the kinematics of tensor fields on the hypersurface,
(III) the dynamics of the fields on the hypersurface,
and (IV) geometrodynamics of the fields\fn{References to these complicated 
papers are pinned down by these Roman numerals followed by the relevant section 
numbers.
We restrict attention to $s = 1$ in this section.}.
The fields are decomposed into perpendicular and tangential parts.
We are mainly concerned with 1-forms in this section, for which the 
decomposition is\fn{We use $e^{\alpha}_a$ for the projector onto the hypersurface and
$n_{\alpha}$ for the perpendicular vector to the hypersurface, and    
$K_{ab}$ for the extrinsic curvature.    
The index perpendicular to the hypersurface is denoted by the 
subscript$_\perp$,
the subscript $_{\not-}$ denotes the
tilt part and the subscript $_{\mbox{\scriptsize t\normalsize}}$ denotes the translational part.}
$A_{\alpha} = n_{\alpha}A_{\perp} + e^a_{\alpha}A_a$;
we also require the decomposition of the metric,
$g_{\alpha\beta} = g_{ab}e^a_{\alpha}e^b_{\beta} - n_{\alpha}n_{\beta}$.
A deformation at a point $x$ of a hypersurface $\Sigma$ may be decomposed 
into two
parts: the \it tilt\normalfont, for which $N(x) = 0$, $[\pa_aN](x) \neq 0$ and 
the \it translation\normalfont, for which $N(x) \neq 0$, $[\pa_aN](x) = 0$.  
We follow Kucha\v{r}'s use of first-order actions.
For the 1-form, this amounts to rewriting the second-order action
$S_{\mbox{\scriptsize A\normalsize}}
= \int d^4x \sqrt{- g} \mbox{\sffamily L\normalfont}
(A_{\alpha}, \nabla_{\beta}A_{\alpha}, g_{\alpha\beta})$
by setting
$\lambda^{\alpha\beta} = \frac{\pa L}{\pa (\nabla_{\beta}A_{\alpha})}$
and using the Legendre transformation
$(A_{\alpha}, \nabla_{\beta}A_{\alpha}, \mbox{\sffamily L\normalfont}) \longrightarrow (A_{\alpha}, \lambda_{\alpha\beta}, L)$,
where the `Lagrangian potential' is 
$L = [\lambda^{\alpha\beta}\nabla_{\beta}A_{\alpha} - \mbox{\sffamily L\normalfont}](A_{\alpha}, \lambda_{\alpha\beta}, g_{\alpha\beta})$.
Then the `hypersurface Lagrangian' is
\be
{\delta_N}   S_{\mbox{\scriptsize A\normalsize}}
= \int_{\Sigma}\mbox{d}^3x(\pi^{\perp}\delta_{\mbox{\bf\scriptsize 
N\normalsize\normalfont}}A_{\perp}
+ \pi^a\delta_{\mbox{\bf\scriptsize N\normalsize\normalfont}}A_a -
N _{\mbox{\scriptsize A\normalsize}}{\cal H}^{\mbox{\scriptsize o\normalsize}}
- N^a {}_{\mbox{\scriptsize A\normalsize}}{\cal H}^{\mbox{\scriptsize o\normalsize}}_a)
\label{covectoraction}
\ee
where $\delta_N$ is the normal change in the projection, the A-contribution to the momentum constraint
$_{\mbox{\scriptsize A\normalsize}}{\cal H}^{\mbox{\scriptsize o\normalsize}}_a$ is obtained from 
$\delta_N = \delta_{\mbox{\bf\scriptsize N\normalsize\normalfont}} - \pounds_{\xi}$ (see Fig. 1)
integrating by parts where necessary, and the A-contribution to the Hamiltonian constraint on a fixed background 
$_{\mbox{\scriptsize A\normalsize}}{\cal H}^{\mbox{\scriptsize o\normalsize}}$ may be further decomposed into its 
translation and tilt parts,
\be
_{\mbox{\scriptsize A\normalsize}}{\cal H}^{\mbox{\scriptsize o\normalsize}}  =
{}_{\mbox{\scriptsize A\normalsize}}{\cal H}^{\mbox{\scriptsize o\normalsize}}_{\mbox{\scriptsize t\normalsize}}  
+ {_{\mbox{\scriptsize A\normalsize}}{\cal H}^{\mbox{\scriptsize o\normalsize}}_{\not-}}.
\label{hamdec}
\ee
The translational part $_{\mbox{\scriptsize A\normalsize}}{\cal H}^{\mbox{\scriptsize o\normalsize}}_{\mbox{\scriptsize t\normalsize}}$ may 
contain a term $2 {}_{\mbox{\scriptsize A\normalsize}}P^{ab}K_{ab}$ due to the possibility of \it derivative coupling \normalfont of the metric 
to the 1-form, whilst the remainder of $_{\mbox{\scriptsize A\normalsize}}{\cal H}^{\mbox{\scriptsize o\normalsize}}_{\mbox{\scriptsize t\normalsize}}$ 
is denoted by $_{\mbox{\scriptsize A\normalsize}}{\cal H}_{\mbox{\scriptsize t\normalsize}}$: 
\be
_{\mbox{\scriptsize A\normalsize}}{\cal H}^{\mbox{\scriptsize o\normalsize}}_{\mbox{\scriptsize t\normalsize}}  =
{}_{\mbox{\scriptsize A\normalsize}}{\cal H}_{\mbox{\scriptsize t\normalsize}} + 2 {}_{\mbox{\scriptsize 
A\normalsize}}P^{ab}K_{ab}.
\label{transdec}
\ee

For the 1-form field, using the decomposition
$\lambda^{\alpha\beta} = \lambda^{\perp\perp}n^{\alpha}n^{\beta} + 
\lambda^{a\perp}e^{\alpha}_a n^{\beta}
+ \lambda^{\perp b}n^{\alpha}e^{\beta}_b + 
\lambda^{ab}e^{\alpha}_ae^{\beta}_b$
and $\lambda^{a\perp} = \pi^a$, $\lambda^{\perp\perp} = \pi^{\perp}$
(by the definition of canonical momentum), one obtains
\be
_{\mbox{\scriptsize A\normalsize}}{\cal H}_{\mbox{\scriptsize t\normalsize}} = L + \sqrt{h}
(\lambda^{\perp a}D_a A_{\perp} - \lambda^{ab}D_aA_b).
\label{1formtrans}
\ee
We also require
\be
_{\mbox{\scriptsize A\normalsize}}P^{ab} = \frac{\sqrt{h}}{2}( - 
A^{(a|}\lambda^{\perp|b)}
+ A_{\perp}\lambda^{(ab)} - A^{(a}\pi^{b)}).
\label{1formderiv}
\ee
For the 1-form, $\lambda^{\perp a}$ and $\lambda^{ab}$ play the role of 
Lagrange multipliers;
one would then use the corresponding multiplier equations to attempt
to eliminate the multipliers from (\ref{covectoraction}).  In our examples 
below, $A_{\perp}$
will also occur as a multiplier, but this is generally not the case.

The above sort of decomposition holds for any rank of tensor field.   
${\cal H}^{\mbox{\scriptsize o\normalsize}}_{\not-}$, $P^{ab}$ and $\pounds_{\xi}$ are universal for each rank, 
whereas ${\cal H}_{\mbox{\scriptsize t\normalsize}}$ contains $L$, which has further details of the particular field in question.  
These three universal features represent the kinematics due to the presupposition of spacetime.  
The $\pounds_{\xi}$ contribution is `shift kinematics', whilst the tilt contribution is `lapse kinematics'.  

The point of Kucha\v{r}'s papers is to construct very general consistent matter theories by 
presupposing spacetime and correctly implementing the resulting kinematics.    
We are able to show below that in not presupposing spacetime, BF\'{O} are attempting to construct 
consistent theories by using shift kinematics (which is the best matching principle) alone, and thus attempting to deny the  
presence of any `lapse kinematics' in nature.  This turns out to be remarkably successful for the bosonic theories of nature.     

We begin by noting that nonderivative-coupled fields are a lot simpler to deal with than 
derivative-coupled ones.  We then ask which fields are included in this simpler case, 
in which the matter fields do not affect the gravitational part of the Hamiltonian constraint  
so that the gravitational momenta remain independent of the matter fields.  
Now, we realize that this is a \it tacit assumption \normalfont in almost all\fn{We have argued in 
Sec II.E that the exception, Brans--Dicke theory, is a mild one.} of BF\'{O}'s work.

\noindent \bf 0 \normalfont : the implementation of `adding on' matter is for 
matter contributions that do not interfere with the structure of the gravitational theory.

\noindent This amounts to the absence of Christoffel symbols in the matter Lagrangians, which is true of 
minimally-coupled scalar fields ($D_a\chi = \pa_a\chi$) and of Maxwell and Yang--Mills theories and their massive counterparts 
(since $D_aA_b - D_bA_a = \pa_aA_b - \pa_bA_a$).  Thus it suffices to start off by considering the 
nonderivative-coupled case on the grounds that it includes all the fields hitherto
thought to fit in with the BF\'{O} scheme, and also the massive 1-form fields which do not.  

Consider then the Proca 1-form.  Its Lagrangian is
\be
\mbox{\sffamily L\normalfont}_{\mbox{\scriptsize Proca\normalsize}}
= - \nabla_{[\alpha}A_{\beta]}\nabla^{[\alpha}A^{\beta]} - 
\frac{m^2}{2}A_{\alpha}A^{\alpha},
\label{lagproca}
\ee
with corresponding Lagrangian potential
\be
L = -\frac{1}{4}\lambda^{[\alpha\beta]}\lambda_{[\alpha\beta]} + 
\frac{m^2}{2}A_{\alpha}A^{\alpha}.
\ee
Whereas $_{\mbox{\scriptsize A\normalsize}}{\cal H}_{\not-}^{\mbox{\scriptsize o\normalsize}}$ has in fact 
been completed to a divergence,
$_{\mbox{\scriptsize (A)\normalsize}}{\cal H}^{\mbox{\scriptsize o\normalsize}}_{\not-} = A^aD_a\pi^{\perp} 
+ A_{\perp} D_a\pi^a$
suffices to generate the tilt change of $A_{\perp}$ and $A_a$ for the 
universal 1-form (see III.6).
The first term of this vanishes since $\pi^{\perp} = 0$ by antisymmetry for the 1-forms described by 
(\ref{lagproca}).
Also $_{\mbox{\scriptsize A\normalsize}}P^{ab} = 0$ by antisymmetry so
$$
_{\mbox{\scriptsize A\normalsize}}{\cal H}^{\mbox{\scriptsize o\normalsize}} =
\sqrt{h}\left[ -\frac{1}{4}\lambda_{ab}\lambda^{ab} + \frac{1}{2h}\pi_a\pi^a \right.
$$
\be
\left.
+ \frac{m^2}{2}(A_aA^a - A_{\perp}^2) 
- \lambda^{ab}A_{[a,b]} \right] + 
A_{\perp}D_a\pi^a
\label{Procaham}
\ee
by (\ref{hamdec}, \ref{transdec}).  
The multiplier equation for $\lambda_{ab}$ gives 
\be
\lambda_{ab} = -2 D_{[b}A_{a]} \equiv B_{ab}.
\label{lme}
\ee
For $m \neq 0$, the multiplier equation for $A_{\perp}$ gives
\be
A_{\perp} = - \frac{1}{m^2\sqrt{h}} D_a\pi^a,
\label{lightening}
\ee
and elimination of the multipliers in (\ref{Procaham}) using (\ref{lme}, \ref{lightening}) gives
$$
_{\mbox{\scriptsize A\normalsize}}{\cal H}^{\mbox{\scriptsize o\normalsize}} = \frac{1}{2\sqrt{h}}\pi_a\pi^a 
+ \frac{\sqrt{h}}{4}B_{ab}B^{ab}
+ \frac{m^2\sqrt{h}}{2}A_aA^a 
$$
\be
+ \frac{1}{2m^2\sqrt{h}}(D_a\pi^a)^2,
\label{unel}
\ee
which is non-ultralocal in the momenta.
We note that this does nothing to eliminate the remaining term in the tilt:
the Proca field has nonzero tilt.  

But, for $m = 0$, the $A_{\perp}$ multiplier equation
gives instead the Gauss constraint of electromagnetism 
\be
{\cal G} \equiv D_a\pi^a \approx 0.
\label{Gaussweak}
\ee
This would not usually permit $A_{\perp}$ to be eliminated from 
(\ref{unel}) but the final form of $_{\mbox{\scriptsize A\normalsize}}{\cal H}^{\mbox{\scriptsize o\normalsize}}$ for $m = 0$ is 
\be
_{\mbox{\scriptsize A\normalsize}}{\cal H}^{\mbox{\scriptsize o\normalsize}} = \frac{\sqrt{h}}{4}B^{ab}B_{ab} + 
\frac{1}{2\sqrt{h}}\pi_a\pi^a + A_{\perp}(D_a\pi^a \approx 0), 
\ee
so the cofactor of $A_{\perp}$ in (\ref{covectoraction}) weakly vanishes by $\ref{Gaussweak}$, 
so $A_{\perp}$ may be taken to `accidentally' drop out.
This means that the tilt of the Maxwell field may be taken to be zero.
The tilt is also zero for the metric and for the scalar field.  
So far all these fields are allowed by BF\'{O} and have no tilt, 
whereas the disallowed Proca field has tilt.  

We can begin to relate this occurrence to the BSW principle \bf 2 \normalfont or \bf 2G\normalfont.    
For, suppose an action has a piece depending on $\pa_aN$ in it.  Then the immediate elimination of 
$N$ from it is \it not \normalfont algebraic, so the procedure of BSW is not possible.  
By definition, the tilt part of the Hamiltonian constraint is built from the $\pa_aN$ 
contribution using integration by parts.  
But, for the $A_{\perp}$-eliminated Proca Lagrangian,
this integration by parts gives a term that is
non-ultralocal in the momenta, $(D_a\pi^a)^2$,
which again contain $\pa_aN$ within.
Thus, for this formulation of Proca theory, one cannot build a BSW--Proca action to start off with.   
Of importance, this problem with spatial derivatives
was not foreseen in the simple analogy with the Jacobi principle in mechanics,
where there is only one independent variable.  

The above argument requires refinement from the treatment of further important physical examples.  
This is a fast method of finding matter theories compatible with the 3-space approach by the following argument.  
If there is no derivative coupling and if one can arrange for the tilt to play no part in a formulation of a matter theory, 
then all that is left of the hypersurface kinematics is the shift kinematics, which is the best-matching principle.  
But complying with hypersurface kinematics is a guarantee for consistency so in these cases best matching suffices for consistency.

First, we consider K interacting 1-forms
$A_a^{\mbox{\bf\scriptsize K\normalsize\normalfont}}$
with Lagrangian\fn{$\bf D \normalfont_a$ is the Yang--Mills covariant derivative and 
$C_{{\mbox{\scriptsize\bf ABC\normalsize\normalfont}}}$  are the Yang--Mills 
structure constants.  By $\mbox{\sffamily g\normalfont}
{C^{\mbox{\bf\scriptsize A\normalsize\normalfont}}}_{\mbox{\bf\scriptsize 
BC\normalsize\normalfont}}$ we strictly mean $\mbox{\sffamily g\normalfont}^{\cal A}
C^{\mbox{\bf\scriptsize A\normalsize\normalfont}}_{{\cal A}\mbox{\bf\scriptsize BC\normalsize\normalfont}}$ where ${\cal A}$ indexes each gauge subgroup in a direct product.  
Then each such gauge subgroup can be associated with a distinct coupling constant $\mbox{\sffamily g\normalfont}^{\cal A}$.}.  
$$
\mbox{\sffamily L\normalfont}^{\mbox{\scriptsize ${\cal A}$\normalsize}}
= - \left(\nabla_{[\alpha}A^{\mbox{\bf\scriptsize 
A\normalsize\normalfont}}_{\beta]}
+ \frac{\mbox{\sffamily g\normalfont}}{2}
{C^{\mbox{\bf\scriptsize A\normalsize\normalfont}}}_{\mbox{\bf\scriptsize 
BC\normalsize\normalfont}}
A^{\mbox{\bf\scriptsize B\normalsize\normalfont}}_{\beta}
A^{\mbox{\bf\scriptsize C\normalsize\normalfont}}_{\alpha}\right)\times 
$$
\be
\left(
\nabla^{[\alpha}A_{\mbox{\bf\scriptsize A\normalsize\normalfont}}^{\beta]}
+ \frac{\mbox{\sffamily g\normalfont}}{2}
{C_{\mbox{\bf\scriptsize ADE\normalsize\normalfont}}}A^{\mbox{\bf\scriptsize 
D\normalsize\normalfont}\beta}
A^{\mbox{\bf\scriptsize E\normalsize\normalfont}\alpha}
\right)
- \frac{m^2}{2}A_{\alpha{\mbox{\bf\scriptsize M\normalsize\normalfont}}}
A^{\alpha{\mbox{\bf\scriptsize M\normalsize\normalfont}}}.
\label{lagYMmass}
\ee
We define 
$\lambda^{\alpha\beta}_{\mbox{\bf\scriptsize M\normalsize\normalfont}}
= \frac{\pa \mbox{\sffamily \scriptsize L\normalsize\normalfont}}
{\pa (\nabla_{\beta}A^{\mbox{\bf\scriptsize 
M\normalsize\normalfont}}_{\alpha})}$ 
and the corresponding Lagrangian potential is
\be
L = -\frac{1}{4}\lambda^{[\alpha\beta]}_{\mbox{\bf\scriptsize 
M\normalsize\normalfont}}
\lambda^{\mbox{\bf\scriptsize M\normalsize\normalfont}}_{[\alpha\beta]}
- \frac{\mbox{\sffamily g\normalfont}}{2}
C_{\mbox{\bf\scriptsize BDE\normalsize\normalfont}}
A_{\beta}^{\mbox{\bf\scriptsize D\normalsize\normalfont}}
A_{\alpha}^{\mbox{\bf\scriptsize E\normalsize\normalfont}}
\lambda^{\alpha\beta{\mbox{\bf\scriptsize B\normalsize\normalfont}}}
+ \frac{m^2}{2}A_{\alpha{\mbox{\bf\scriptsize M\normalsize\normalfont}}}
A^{\alpha{\mbox{\bf\scriptsize M\normalsize\normalfont}}}.
\ee
The overall tilt contribution is now the sum of the tilt contributions of the individual fields, so  
$_{(A_{\mbox{\bf\scriptsize M\normalsize\normalfont}})}{\cal H}^{\mbox{\scriptsize o\normalsize}}_{\not-}
= A^{\mbox{\bf\scriptsize M\normalsize\normalfont}}_{\perp}D_a
\pi^a_{\mbox{\bf\scriptsize M\normalsize\normalfont}}$
suffices to
generate the
tilt change.
Again,
$_{{\mbox{\scriptsize A\normalsize}}_{\mbox{\bf\scriptsize 
M\normalsize\normalfont}}}P^{ab} = 0$
by antisymmetry so 
$$
_{\mbox{\scriptsize A\normalsize}_{\mbox{\bf\scriptsize 
M\normalsize\normalfont}}}{\cal H}^{\mbox{\scriptsize o\normalsize}}
= \sqrt{h}
\left[ 
-\frac{1}{4}\lambda^{\mbox{\bf\scriptsize M\normalsize\normalfont}}_{ab}\lambda_{\mbox{\bf\scriptsize M\normalsize\normalfont}}^{ab}
+ \frac{1}{2h}\pi^{\mbox{\bf\scriptsize M\normalsize\normalfont}}_a \pi_{\mbox{\bf\scriptsize M\normalsize\normalfont}}^a
\right.
$$
$$
\left.
+ \frac{m^2}{2}(A^{\mbox{\bf\scriptsize M\normalsize\normalfont}}_a
A_{\mbox{\bf\scriptsize M\normalsize\normalfont}}^a -
A^{\mbox{\bf\scriptsize M\normalsize\normalfont}}_{\perp}
A_{\perp{\mbox{\bf\scriptsize M\normalsize\normalfont}}}) -
\lambda_{\mbox{\bf\scriptsize M\normalsize\normalfont}}^{ab}
A^{\mbox{\bf\scriptsize M\normalsize\normalfont}}_{[a,b]}\right]  +
A^{\mbox{\bf\scriptsize M\normalsize\normalfont}}_{\perp}{\mbox{\bf 
D\normalfont}}_a
\pi_{\mbox{\bf\scriptsize M\normalsize\normalfont}}^a 
$$
\be
-
\frac{\mbox{\sffamily g\normalfont}}{2}C_{\mbox{\bf\scriptsize 
MPQ\normalsize\normalfont}}(
\sqrt{h}\lambda^{ab{\mbox{\bf\scriptsize M\normalsize\normalfont}}}
A^{\mbox{\bf\scriptsize P\normalsize\normalfont}}_bA^{\mbox{\bf\scriptsize 
Q\normalsize\normalfont}}_a 
+2 \pi^{\mbox{\bf\scriptsize M\normalsize\normalfont}}
A^{\mbox{\bf\scriptsize P\normalsize\normalfont}}_{\perp}
A^{\mbox{\bf\scriptsize Q\normalsize\normalfont}}_a)
\ee
by (\ref{hamdec}, \ref{transdec}).  
The multipliers are
$\lambda^{ab}_{\mbox{\bf\scriptsize M\normalsize\normalfont}}$
and $A_\perp^{\mbox{\bf\scriptsize M\normalsize\normalfont}}$,
with corresponding
multiplier equations
\be
\lambda^{\mbox{\bf\scriptsize M\normalsize\normalfont}}_{ab}
= -2 D_{[b}A^{\mbox{\bf\scriptsize M\normalsize\normalfont}}_{a]}
\equiv B^{\mbox{\bf\scriptsize M\normalsize\normalfont}}_{ab},
\ee
$$
A_{\mbox{\bf\scriptsize M\normalsize\normalfont}\perp} = - 
\frac{1}{m^2\sqrt{h}}
\mbox{\bf D\normalfont}_a\pi^a_{\mbox{\bf\scriptsize 
M\normalsize\normalfont}}
$$
\be
\equiv - \frac{1}{m^2\sqrt{h}}(D_a\pi^a{\mbox{\bf\scriptsize 
M\normalsize\normalfont}}
+ \mbox{\sffamily g\normalfont}C_{\mbox{\bf\scriptsize 
LMP\normalsize\normalfont}}
\pi^{\mbox{\bf\scriptsize L\normalsize\normalfont}a}
A_a^{\mbox{\bf\scriptsize P\normalsize\normalfont}})
\ee
for $m \neq 0$.
We thus obtain the eliminated form
$$
_{{\mbox{\scriptsize A\normalsize}}_{\mbox{\bf\scriptsize 
M\normalsize\normalfont}}}{\cal H}^{\mbox{\scriptsize o\normalsize}}
= \frac{1}{2\sqrt{h}}
\pi_{\mbox{\bf\scriptsize 
M\normalsize\normalfont}a}\pi^{\mbox{\bf\scriptsize 
M\normalsize\normalfont}a}
+ \frac{\sqrt{h}}{4}B_{\mbox{\bf\scriptsize M\normalsize\normalfont}ab}
B^{\mbox{\bf\scriptsize M\normalsize\normalfont}ab}
$$
\be
+ \frac{m^2\sqrt{h}}{2}
A_{\mbox{\bf\scriptsize M\normalsize\normalfont}a}A^{\mbox{\bf\scriptsize 
M\normalsize\normalfont}a}
+ \frac{1}{2m^2\sqrt{h}}
(\mbox{\bf D\normalfont}_a\pi^{\mbox{\bf\scriptsize 
M\normalsize\normalfont}a})
(\mbox{\bf D\normalfont}_b\pi_{\mbox{\bf\scriptsize 
M\normalsize\normalfont}b})
\ee
and the massive Yang--Mills field is left with nonzero tilt.
For $m = 0$, the second multiplier equation gives instead the Yang--Mills 
Gauss constraint
\be
{\cal G}^{\mbox{\bf\scriptsize M\normalsize\normalfont}}
\equiv {\mbox{\bf D\normalfont}}_a\pi^{\mbox{\bf\scriptsize 
M\normalsize\normalfont}a} \approx 0.
\label{YMGaussweak}
\ee
In this case, the tilt is nonzero, but the Yang--Mills Gauss constraint 
`accidentally' enables the derivative part of the tilt to be converted into 
an algebraic expression, which then happens to cancel with part of the Lagrangian potential: 
$$
_{{\mbox{\scriptsize A\normalsize}}_{\mbox{\bf\scriptsize M\normalsize\normalfont}}}{\cal H}^{\mbox{\scriptsize o\normalsize}} 
= \frac{\sqrt{h}}{4}
B^{ab}_{\mbox{\bf\scriptsize M\normalsize\normalfont}}
B_{ab}^{\mbox{\bf\scriptsize M\normalsize\normalfont}} 
+ \frac{1}{2\sqrt{h}}
\pi^{\mbox{\bf\scriptsize M\normalsize\normalfont}}_a
\pi^a_{\mbox{\bf\scriptsize M\normalsize\normalfont}} 
$$
\be
+ A^{\mbox{\bf\scriptsize M\normalsize\normalfont}}_{\perp}
(D_a\pi^a_{\mbox{\bf\scriptsize M\normalsize\normalfont}} 
+ \mbox{\sffamily g\normalfont}
C_{\mbox{\bf\scriptsize LMP\normalsize\normalfont}}
\pi^{\mbox{\bf\scriptsize L\normalsize\normalfont}a}
A_a^{\mbox{\bf\scriptsize P\normalsize\normalfont}} \approx 0).
\ee

Second, we consider $U(1)$ 1-form--scalar gauge theory, with interactions of the form 
$\chi^*A^{\mu}\pa_{\mu}\chi$ and $\chi^*\chi A^{\mu}A_{\mu}$.  This could be viewed either as the interaction of a 
(strongly favoured but still hypothetical) Higgs field with the electromagnetic field, or as 
a warm-up exercise toward the inclusion of the interaction term of Maxwell--Dirac theory 
[the classical theory behind quantum electrodynamics (QED)] 
and its Standard Model generalization (see Sec. V.B).
The Maxwell--scalar Lagrangian is\fn{This working is unaffected by inclusion of a scalar field potential function.}
$$
\mbox{\sffamily L\normalfont}^{\mbox{\scriptsize U\normalsize}(1)}_{\mbox{\scriptsize MS\normalsize}}
= - \nabla_{[\alpha}A_{\beta]}\nabla^{[\alpha}A^{\beta]} +
(\pa_{\mu}\chi - ieA_{\mu}\chi)(\pa^{\mu}\chi^* + ieA^{\mu}\chi^*) 
$$
\be
- \frac{m_{\chi}^2}{2}\chi^*\chi.
\label{lagU(1)}
\ee
Now, in addition to $\lambda^{\alpha\beta}$, define
$\mu^{\alpha} = \frac{\pa\mbox{\sffamily\scriptsize 
L\normalsize\normalfont}}{\pa (\nabla_{\alpha}\chi)}$ and
$\nu^{\alpha} = \frac{\pa\mbox{\sffamily\scriptsize 
L\normalsize\normalfont}}{\pa (\nabla_{\alpha}\chi^*)}$,
so the Lagrangian potential is 
$$
L = -\frac{1}{4}\lambda^{[\alpha\beta]}\lambda_{[\alpha\beta]} + 
\frac{m^2}{2}A_{\alpha}A^{\alpha}
+ \mu^{\alpha}\nu_{\alpha} - ieA_{\alpha}(\chi^*\nu^{\alpha} - 
\chi\mu^{\alpha}) 
$$
\be
+ \frac{m_{\chi}^2}{2}\chi^*\chi.
\ee
$_{\mbox{\scriptsize (A)\normalsize}}{\cal H}^{\mbox{\scriptsize o\normalsize}}_{\not-} =   A_{\perp} 
D_a\pi^a$ still
suffices to generate the tilt (as scalars contribute no tilt), we have
$_{A,\chi,\chi^*}P^{ab} = 0$,
and
$$
_{A,\chi,\chi^*}{\cal H}^{\mbox{\scriptsize o\normalsize}}_{\mbox{\scriptsize t\normalsize}} =  \sqrt{h}
\left[
- \frac{1}{4}\lambda_{ab}\lambda^{ab} + \mu_a\nu^a \right. 
$$
$$
\left. 
+ \frac{1}{h}
\left(
\frac{1}{2}\pi_a\pi^a 
- \pi_{\chi}\pi_{\chi^*}
\right) 
+ \frac{m_{\chi}^2}{2}\chi^*\chi 
\right.
$$
\be
\left.
- ie\left( A_a[\chi^*\nu^a - \chi\mu^a] 
- \frac{A_{\perp}}{\sqrt{h}}[\chi^*\pi_{\chi^*} - \chi\pi_{\chi}]\right)   
\right].
\label{unelUH}
\ee
The $\lambda_{ab}$ multiplier equation is (\ref{lme}) again, whilst the $A_{\perp}$ multiplier equation is now 
\be
{\cal G}_{\mbox{\scriptsize U\normalsize\normalfont}(1)}
\equiv D_a\pi^a + ie(\chi^*\pi_{\chi^*} - \chi\pi_{\chi}) = 0,
\label{sourceGauss}
\ee
which can be explained in terms of electromagnetism now having a fundamental source.  
In constructing $_{A,\chi,\chi^*}{\cal H}^{\mbox{\scriptsize o\normalsize}} $ from (\ref{hamdec}, \ref{transdec}, \ref{unelUH}), 
we can convert the tilt to an algebraic expression by the sourced Gauss law (\ref{sourceGauss}) 
which again happens to cancel with a contribution from the Lagrangian potential: 
$$
_{A,\chi,\chi^*}{\cal H}^{\mbox{\scriptsize o\normalsize}} 
= -\lambda^{ab}A_{[a,b]} - (\mu^a + \nu^a)\phi_{,a} + { }_{(A)}{\cal H}^{\mbox{\scriptsize o\normalsize}}_{\not-} 
$$
$$
+ _{A,\chi,\chi^*}{\cal H}^{\mbox{\scriptsize o\normalsize}}_{\mbox{\scriptsize t\normalsize}} 
$$
$$
= 
\left[  
\frac{1}{4}B_{ab}B^{ab} - \mu_a\nu^a 
+ \frac{1}{h}
\left(
\frac{1}{2}\pi_a\pi^a 
- \pi_{\chi}\pi_{\chi^*}
\right) \right.
$$
\be
\left.
+ \frac{m_{\chi}^2}{2}\chi^*\chi
\right]
+ A_{\perp}[D_a\pi^a + ie(\chi^*\pi_{\chi^*} - \chi\pi_{\chi}) \approx 0].
\ee  
It is not too hard to show that the last two accidents also accidentally conspire together to wipe out the 
tilt contribution in Yang--Mills 1-form--scalar gauge theory.
This theory is also obviously nonderivative-coupled.

We now present a more general treatment about the occurrence of these accidents.  
They arise from eliminating $A_{\perp}$ from its multiplier equation.
For this to make sense, $A_{\perp}$ must \it be \normalfont a multiplier,
thus $\pi^{\perp} = 0$.  Then for general $L$, the multiplier equation is
\be
\frac{\pa L}{\pa A_{\perp}} +D_a\pi^a = 0.
\label{multipleq}
\ee
Then the requirement that $A_{\perp}D_a\pi^a + L$ be independent of $A_{\perp}$ on using (\ref{multipleq}) means that  
$- A_{\perp}\frac{\pa L}{\pa A_{\perp}} + L$ is independent of $A_{\perp}$.  Thus the accidents occur whenever the Lagrangian potential is linear in $A_{\perp}$.

From the broadening of our understanding due to the above two examples, we can precisely 
reformulate the BSW principle \bf 2 \normalfont within the GC hypersurface framework as

\noindent \bf 2U \normalfont:
We use lapse-uneliminated actions homogeneously quadratic in their velocities
and permit only those for which the matter contributes a weakly vanishing tilt.

\noindent We can combine this with dropping the requirement for homogeneously 
quadratic actions
(Principle \bf 2G\normalfont) to obtain a Principle \bf 2UG\normalfont, 
in anticipation of the inclusion of spin-$\frac{1}{2}$ fermions.

So for Einstein--Maxwell theory, Einstein--Yang--Mills theory, and their corresponding scalar gauge theories, 
1) the absence of derivative coupling guarantees that they can be coupled to GR without disrupting its canonical structure as tacitly assumed by BF\'{O}.  
2) The absence of tilt guarantees that the resulting coupled theories can be put into BSW form.  
Because the theories have homogeneously quadratic kinetic terms, this is indeed the BSW form \bf 2 \normalfont (as opposed to its generalization \bf 2G\normalfont),
3) now, the GC hypersurface framework guarantees consistency if all the required kinematics are included.  But the only sort of kinematics left is best matching.  
Thus, all these theories are guaranteed to exist as theories in BF\'{O}'s original formulation of the 3-space approach.  

These workings begin to show (if one presupposes spacetime), what sorts of obstacles in Kucha\v{r}'s spacetime ontology 
might be regarded as responsible for the uniqueness results for bosonic matter when one starts from BF\'{O}'s 3-dimensional ontology (see also Sec IV.C).    

There is a slight procedural complication in 3), which we illustrate for the BF\'{O} formulation of Einstein--Maxwell theory.
One starts off with 
$$
S_{\mbox{\scriptsize BSW\normalsize}_{\mbox{\scriptsize A\normalsize}}}
= \int \mbox{d}\lambda \int \mbox{d}^3x \sqrt{h} \sqrt{   R - D_{ [a }A_{ b] }D^{ [a }A^{ b] }   } \times 
$$
\be
\sqrt{T_{\mbox{\scriptsize g\normalsize}} + h^{ab}(\dot{A}_a - 
\pounds_{\xi}A_a)(\dot{A}_b - \pounds_{\xi}A_b)},
\label{ASBSWA}
\ee
and then one discovers the Gauss constraint of electromagnetism ${\cal G}$ is enforced, which one then encodes by the corresponding `electromagnetic' best matching.  
This amounts to the introduction of an auxiliary velocity $\Theta$ (variation w.r.t which yields ${\cal G}$), according to 
\be
\dot{A}_a \longrightarrow \dot{A}_a  - \pa_a\Theta.
\label{correction}
\ee

\subsection{The 3-Space Approach allows more than the Fields of Nature}

We have described how the fields hitherto known to be permitted by the 3-space approach 
may be identified within the GC approach.
These fields all have the universal kinematic feature called best matching by BF\'{O},
and no other significant universal feature (tilt or derivative coupling).
Are these fields then the known fundamental matter fields, which somehow have 
less universal kinematic features than GC would lead one to expect?
This question may be subdivided as follows.  
Does the 3-space approach single out \it only \normalfont the known fundamental matter fields?  
Does the 3-space approach single out \it all \normalfont the known fundamental matter fields?
Kucha\v{r} makes no big deal about the simplified form
weakly equivalent to his decomposition of the electromagnetic field, because it does 
not close to reproduce the Dirac Algebra (see III.11-12); 
it only does so modulo the Gauss constraint of electromagnetism, ${\cal G}$.
He takes this to be an inconvenience, one which can be got round by adhering to 
the form directly obtained from the decomposition, whereas BF\'{O} take it as a virtue that 
the simplified form `points out' the new constraint, ${\cal G}$, as an integrability condition.  

The first question can be answered by counterexample.
One should interpret the question as coarsely as possible;
for example one could argue that the 3-space approach
is not capable of restricting the possibility of 
Yang--Mills theory to the gauge groups conventionally used to describe nature,
or that by no means is massless 1-form--scalar gauge theory guaranteed to occur in nature.
Rather than such subcases or effects due to interaction terms,
we find it more satisfactory to construct a distinct matter theory which is not known 
to be present in nature.  The last subsection has put us into a good position to do this.

Consider the 2-form $\Phi_{\alpha\beta}$ Lagrangian
\be
\mbox{\sffamily L\normalfont} =
- \nabla_{[\gamma}\Phi_{\alpha\beta]}\nabla^{[\gamma}\Phi^{\alpha\beta]}
- \frac{m^2}{2}\Phi_{\alpha\beta}\Phi^{\alpha\beta}, 
\ee
define 
$\lambda^{\alpha\beta\gamma} = \frac{\pa L}{\pa (\pa_{\gamma}\Phi_{\alpha\beta})}$
and use the Legendre transformation to obtain the Lagrangian potential
\be
L = -\frac{1}{4}\lambda^{[\alpha\beta\gamma]}\lambda_{[\alpha\beta\gamma]}
+ \frac{m^2}{2}\Phi_{\alpha\beta}\Phi^{\alpha\beta}.
\ee
Then $_{(\Phi)}{\cal H}^{\mbox{\scriptsize o\normalsize}}_{\not-} = 2\Phi_{\perp b}D_a\pi^{ab}$
suffices to generate the 2-form tilt and $_\Phi P^{ab} = 0$ by antisymmetry.
The multipliers are $\lambda^{abc}$ and $A_{\perp a}$ with corresponding 
multiplier equations $\lambda_{abc} = -2 D_{[b}\Phi_{ab]} \equiv B_{abc}$ and, for $m \neq 
0$,
\be
\Phi^b_{\perp} = - \frac{1}{m^2\sqrt{h}} D_a\pi^{ab},
\ee
which may be used to eliminate the multipliers,
giving rise to the non-ultralocal form
$$
_{\Phi}{\cal H}^{\mbox{\scriptsize o\normalsize}} = \frac{\sqrt{h}}{4}B^{abc}B_{abc} + 
\frac{3}{4\sqrt{h}}\pi^{ab}\pi_{ab}
$$
\be
+\frac{7}{8m^2\sqrt{h}}h_{bd}
(D_a\pi^{ab})(D_c\pi^{cd}) + \frac{m^2}{2}\Phi_{ab}\Phi^{ab}.
\ee

But for $m = 0$, the $\Phi_{\perp b}$ multiplier constraint is 
\be
{\cal G}^b \equiv D_a\pi^{ab} \approx 0
\label{2Gaussweak}
\ee
and
\be
_{\Phi}{\cal H}^{\mbox{\scriptsize o\normalsize}} = \frac{h}{4}B^{abc}B_{abc} + \frac{3}{4\sqrt{h}}\pi^{ab}\pi_{ab} 
+ 2\Phi_{\perp b}(D_a\pi^{ab} \approx 0).
\ee
So our massless 2-form's tilt is zero and this leads to the elimination of $\Phi_{\perp b}$ 
by the same sort of `accident' that permits $A_{\perp}$ to be eliminated in electromagnetism.    
So, for this massless 2-form, best matching is equivalent to all the GC hypersurface kinematics, 
and as this guarantees closure, we deduce that there exists a resulting 3-space approach theory 
starting with
$$
S_{\Phi} = \int \mbox{d}\lambda \int \mbox{d}^3x\sqrt{h}\sqrt{R + D_{[c}\Phi_{ab]}D^{[c}\Phi^{ab]}} \times
$$
\be
\sqrt{T_{\mbox{\scriptsize g\normalsize}} + h^{ab}h^{cd}(\dot{\Phi}_{[ab]} - \pounds_{\xi}\Phi_{ab})(\dot{\Phi}_{[cd]} - \pounds_{\xi}\Phi_{cd}) },
\ee 
which leads to the enforcement of (\ref{2Gaussweak}), which is subsequently encoded by the introduction of an auxiliary variable $\Theta_b$.      
This working should also hold for any $p$-form 
for $p \leq d$, the number of spatial dimensions.  
Yet only the $p = 1$ case, electromagnetism, is known to occur.  
This is evidence against BF\'{O}'s speculation that the 3-space approach hints at `partial unification' of gravity 
and electromagnetism, since these extra unknown fields would also be included as naturally as the electromagnetic field.  
Note also that the ingredients of low energy string theory are
getting included rather than excluded: $p$-forms, the dilatonic coupling...
These are signs that the 3-space approach is not as restrictive as BF\'{O} originally hoped.  

The second question must be answered exhaustively.  It is the minimal requirement for 
the 3-space approach to be taken seriously as a description of nature.  
The 3-space approach gives gravity, electromagnetism and Yang--Mills theories 
such as the $SU(2) \times U(1)$ theory of the electroweak bosons and the $SU(3)$ theory of the 
gluons of the strong force.  One may argue that disallowing fundamental Proca fields is unimportant, 
because the photon and gluons are believed to be massless and the observed masses of the $W^+$, 
$W^-$ and $Z^0$ weak bosons are thought to be not fundamental but rather acquired by spontaneous symmetry breaking \cite{PS}.    
The next problem is the inclusion of spin-$\frac{1}{2}$ fermions (see Sec V.B),  in order to complete the 3-space approach 
for the theories of the simplest free fundamental fields that can account for nature.   
One could then investigate all the interactions involved in the Standard Model \cite{PS}.  
We note that one cannot be sure whether it is these simplest field theories that are present in nature, 
since our particle accelerators are located in a rather flat region.      
Thus our results are subject to our ignorance of nature's unexplored high-curvature regime.  
The notion of `simplest' includes relying on replacing partial derivatives with 
covariant derivatives to find the curved analogues of the flat laws.  Yet this procedure 
could in principle be ambiguous \cite{MTW} or not realized in nature due to putative further symmetry reasons \cite{Kiefer}.  

\subsection{Derivative Coupling and the 3-Space 1-Form Ansatz}

In their study of 1-forms, BF\'{O} used a BSW-type action with potential term 
\be
U_{\mbox{\scriptsize A\normalsize}} = C^{abcd}D_bA_{a}D_dA_{c} + \frac{M^2}{2}A_aA^a,
\label{vfPansatz}
\ee
(where
$C^{abcd} = C_1h^{ac}h^{bd} + C_2h^{ad}h^{bc} + C_3h^{ab}h^{cd}$ for 
constant $C_1$, $C_2$, $C_3$, $M$), which is natural within their 3-space ontology.    
They then obtain $^{\mbox{\scriptsize A\normalsize}}{\cal H}$ and $^{\mbox{\scriptsize A\normalsize}}{\cal H}_i$ in the usual 3-space way 
(from the local square root and from $\xi_i$-variation).  Then the propagation of $^{\mbox{\scriptsize A\normalsize}}{\cal H}$ enforces 
$C_1 = -C_2$ , $C_3 = 0$ and also the Gauss constraint of electromagnetism ${\cal G}$, whose propagation then enforces $M = 0$.  
Having thus discovered that a new (Abelian) gauge symmetry is present, 
${\cal G}$ is then encoded by the corresponding `electromagnetic' best matching,  
by introduction of an auxiliary velocity $\Theta$ [see eq. (\ref{correction})].  
Identifying $\Theta = A_0$, this is a derivation of Einstein--Maxwell theory 
for $A_{\alpha} = [A_0, A_i]$.

We find it profitable to also explain this occurrence starting from the 4-dimensional ontology of the GC hypersurface framework. 
The natural choice of 1-form potential and kinetic terms would then arise from the decomposition of 
\be
\mbox{\sffamily L\normalfont} =
- 
C^{\alpha\beta\gamma\delta}\nabla_{\beta}A_{\alpha}\nabla_{\delta}A_{\gamma}
- \frac{M^2}{2}A_{\alpha}A^{\alpha}.
\label{dostrons}
\ee
Using the following set of four results from (II.2),
\be
\nabla_bA_{\perp} = D_bA_{\perp} - K_{bc}A^c, \\
\label{derivproj1}
\ee
\be
N\nabla_{\perp} A_{a} = - \delta_{N} A_a - NK_{ab}A^b - A_{\perp}\pa_aN \\
\label{derivproj2}
\ee
\be
\nabla_b A_a = D_bA_a - A_{\perp}K_{ab} \\
\label{derivproj3}
\ee
\be
N\nabla_{\perp}A_{\perp} = - \delta_NA_{\perp} - A^a\pa_aN,
\label{derivproj4}
\ee
we obtain that 
$$
\mbox{\sffamily L\normalfont} = 
-(C_1 + C_2 + C_3)
\left(
\frac{\delta_NA_{\perp} + A^a\pa_aN}{N}
\right)^2
$$
$$
+ C_1
\left[
\left(
\frac{ \delta_NA_a + A_{\perp}\pa_aN}{N} + K_{ac}A^c
\right)^2
+ (D_aA_{\perp} - K_{ac}A^c)^2
\right] 
$$
$$
+ 2C_2
\left(
\frac{ \delta_NA_a + A_{\perp}\pa_aN}{N} + K_{ac}A^c
\right)
\left(
D^aA_{\perp} - {K^a}_{c}A^c
\right)
$$
$$
- 2C_3
\left(
\frac{\delta_NA_{\perp} + A^c\pa_cN}{N}
\right)
\left(
D^aA_a - A_{\perp}K
\right) 
$$
$$
- C^{abcd}(D_bA_a - A_{\perp}K_{ab})(D_dA_c - A_{\perp}K_{cd})
$$
\be
- \frac{M^2}{2}(A_aA^a - A_{\perp}A^{\perp}).
\label{starstar}
\ee
Then, if one chooses to prefer the 4-dimensional ontology and then to import  
BF\'{O}'s 3-space assumptions into it, one finds the following explanations for BF\'{O}'s uniqueness results from a 4-dimensional perspective.

First, BF\'{O}'s tacit assumption that addition of a 1-form $A_a$ does not affect the
3-geometry part of the action can be phrased as there being no derivative coupling,
$_{\mbox{\scriptsize A\normalsize}}P^{ab} = 0$, which using 
(\ref{1formderiv})
implies that $\lambda^{(ab)} = 0$, $\pi^b = -\lambda^{\perp b}$.
Since $\lambda^{\alpha\beta} = 
-2C^{\alpha\beta\gamma\delta}\nabla_{\delta}A_{\gamma}$, this \it by itself \normalfont implies $C_1 = - C_2$, $C_3 = 0$.

If $A_{\perp}$ were a velocity as Barbour would argue \cite{partI} (following from its auxiliary status, just as $N$ and $\xi_i$ are velocities),
it makes sense for the 3-space ansatz to contain no $\delta_NA^{\perp}$.
But we now see from (\ref{starstar}) that this by itself is also equivalent to
$C_1 = - C_2$, $C_3 = 0$ from the 4-dimensional perspective.  
Also, inspecting (\ref{starstar}) for Maxwell theory reveals that
\be
\mbox{\sffamily L\normalfont} = \frac{C_1}{N^2}[\delta_NA_a - 
D_a(-NA_{\perp})]^2 - C_1D^bA^a(D_bA_a - D_aA_b).
\ee
So in fact $\Theta = - NA_{\perp}$, so $A_{\perp}$ itself is not a velocity.
Notice in contrast that the issue of precisely what $\Theta$ is does not arise in the 3-space approach because 
it is merely an auxiliary velocity that appears in the last step of the working.

One argument for the 3-space 1-form field ansatz is simplicity: consideration of a 
3-geometry and a single 3-d 1-form leads to Maxwell's equations.   However, we
argue that in the lapse uneliminated form, provided that one is willing to accept the additional 
kinematics,  we can extend these degrees of freedom to include a
dynamical $A_{\perp}$.   The 3-space approach is about \it not \normalfont
accepting kinematics other than best matching,  but the GC hypersurface framework enables us to explore what
happens when tilt and derivative-coupling kinematics are `switched on'. Working within the GC hypersurface framework, 
if $A_{\perp}$ is allowed to be dynamical, there is derivative coupling,  and consistency would require the 
presence of 2 further bunches of terms, with coefficients  proportional to 
$C_1 +  C_2$ and to $C_3$. The first bunch consists of the following sorts of 
terms:   
$$
D^bA^aA_{\perp}\delta_{N}h_{ab}\mbox{ , }
A^b\left(D^aA_{\perp} - A_{\perp}\frac{\pa^aN}{N}\right)\delta_Nh_{ab} 
\mbox{ , }
$$
$$
\frac{1}{N}h^{ab}A^c\delta_NA_a\delta_Nh_{bc}\mbox{ , } 
$$
\be
A^bA^dh^{ac}\delta_{N}h_{ab}\delta_{N}h_{cd}\mbox{ ,  }
A_{\perp}A_{\perp}h^{ac}h^{bd}\delta_Nh_{ab}\delta_Nh_{cd}.
\label{firstbunch}
\ee
The second bunch consists of the following sorts of terms:
$$
h^{ab}\left(A_{\perp}D^cA_c + A^c\frac{\pa_cN}{N}\right)\delta_Nh_{ab} \mbox{ , }
\frac{1}{N}h^{ab}A_{\perp}\delta_NA_{\perp}\delta_Nh_{ab} \mbox { , }
$$
\be
A_{\perp}A_{\perp}h^{ab}h^{cd}\delta_Nh_{ab}\delta_Nh_{cd}.
\label{secondbunch}
\ee
The naive blockwise Riemannian structure of the configuration space of
GR and nonderivative-coupled bosonic fields (\ref{min}) can get badly
broken by derivative coupling (c.f IV.5).
Either of the above bunches by itself exhibits all the unpleasant 
configuration space features we mentioned in Sec II.E:
the first two terms of (\ref{firstbunch}) are linear and hence the
geometry is not Riemannian,  the third is a metric-matter cross-term,
and the last two terms breach the DeWitt structure;
likewise the first term of (\ref{secondbunch}) is linear,
the second is a cross-term and the third is a breach of the DeWitt structure.  
If the DeWitt structure is breached in nature, then the study of pure canonical 
gravity and of the isolated configuration space of pure gravity are undermined.  
Whereas there is no evidence for this occurrence, we have argued at the end of 
the last subsection that some forms of derivative coupling are only manifest 
in experimentally-unexplored high-curvature regimes.  

In the hypersurface framework, if $A_{\perp}$ were dynamical,
then it would not be a Lagrange multiplier, and so it would not have a
corresponding multiplier equation with which the tilt could be
`accidentally' removed, in which case there would not exist a
corresponding BSW form containing $A_{\perp}$.  This argument however is not
watertight, because it does not prevent some other BSW form from existing
since variables other than $A_{\perp}$ could be used in attempts to write down 
actions that obey the 3-space principles.  As an example of such an attempt, 
we could use the $N$-dependent variable $A_0$ to put Proca theory into BSW form.  
In this case the attempt fails as far as the 3-space approach is concerned, because $A_0$ features as a 
non-best-matched velocity in contradiction with principle \bf 1\normalfont.  
This shows however that criteria for whether a matter theory can be coupled to GR 
in the 3-space approach are unfortunately rather dependent on the formalism 
used for the matter field.  The 3-space approach would then amount to attaching 
particular significance to formalisms meeting its description.  This is similar 
in spirit to how those formalisms which close precisely as the Dirac Algebra are 
favoured in the hypersurface framework and the HKT and Teitelboim \cite{Teitelboim} papers.    
In both cases one is required to find at least one compatible formalism for all 
the known fundamental matter fields.  

\section{Discussion and the Inclusion of Spin-$\frac{1}{2}$ Fermions}

\subsection{Variations on the Seventh Route to Relativity}

The split (\ref{hamdec}, \ref{transdec}) of $_{\mbox{\scriptsize A\normalsize}}{\cal H}^{\mbox{\scriptsize o\normalsize}}$
or perhaps more simply the equations (\ref{derivproj1}, \ref{derivproj2},
\ref{derivproj3}, \ref{derivproj4})
(and their analogues for higher-rank tensors [see e.g (III.9)]),
sum up the position of best matching within the GC hypersurface framework.
The required presupposition of embeddability in the GC hypersurface framework leads to three 
sorts of kinematics for tensor fields:  best matching, tilt and derivative coupling. 
All three of these are required in general in order to guarantee consistency and 
Kucha\v{r}'s papers are a recipe for the computation of all the terms required for 
this consistency.  Thus in GR where it is available, the GC hypersurface framework is powerful and 
advantageous as a means of writing down consistent matter theories.
If conformal gravity is regarded as a competing theory to GR, it makes sense therefore 
to question what the 4-geometry of conformal gravity is,  
and whether its use could lead to a more illuminating understanding of matter coupling 
than offered by the 3-space approach.  We are thus free to ask how special GR is 
in admitting a constructive kinematic scheme for coupled consistent tensorial 
matter theories.

As BF\'{O} formulate it, the 3-space approach denies the primary existence of 
the lapse.  But we have demonstrated that whether or not 
the lapse is eliminated does not affect the mathematics, 
so we would prefer to think of the 3-space approach as denying `lapse kinematics'.
BF\'{O}'s use of BSW forms does lead to a more restrictive scheme than GC, but 
we have demonstrated in Sec IV that this restriction can be understood in terms of 
when the GC hypersurface framework has no tilt.
Furthermore, we have unearthed the tacit simplicity postulate \bf 0 \normalfont 
and have rephrased this and the generalized BSW postulate \bf 2G \normalfont 
as nonderivative coupling and the no tilt condition \bf 2UG \normalfont 
respectively within the GC hypersurface framework.   

Working in the GC hypersurface framework (with lapse-uneliminated actions with only shift kinematics) 
has the additional advantage that we are immediately able to turn on and hence 
investigate the mathematical and physical implications of the tilt and derivative-coupling kinematics.  
Nevertheless, it is striking that best matching kinematics suffice to describe all of 
the known fundamental bosonic fields coupled to GR.  
The absence other kinematics includes the absence of the derivative-coupled 
theories whose presence in nature would undermine the study of pure canonical gravity of DeWitt and others.  
We see our work as support for this study.  
The less structure is assumed in theoretical physics, the more room is left for 
predictability.  Could it really be that nature has less kinematics than the GC hypersurface 
framework of GR might have us believe?  

We next question whether the best-matching kinematics itself should be presupposed,
since it is also striking that the additional constraints of the  GR-boson 
system (${\cal H}_i$, ${\cal G}$, ${\cal G}^J$, ...) are interpretable 
as integrability conditions for ${\cal H}$.  This allows the following
alternative to starting with the best-matching principle
\bf 1\normalfont, which could in principle allow more
complicated shift kinematics than the current formulation.\footnote{We 
consider the difference between shift kinematics and lapse kinematics 
to be particularly significant because of their association with linear and quadratic 
constraints respectively.  We have no doubt in the correctness of handling
linear constraints in physics so it would not be a problem if the concept of 
best matching requires refinement.}

\noindent \bf 1I \normalfont :  start with a 3-dimensional action with
bare velocities.  ${\cal H}$ can be deduced immediately from the action, and demanding 
$\dot{{\cal H}} \approx 0$ leads to a number of other constraints.  These are all
then to be encoded by use of auxiliary variables.

\noindent  This has the immediate advantage of treating the
gravitational best matching on the same footing as the encoding of
Gauss constraints.  The 3-space approach has recently
been reformulated this way by \'{O} Murchadha \cite{omurchadha}.

We present caveats to this approach both here and in Sec V.B.  
Here, we note that for strong gravity, \bf 1 \normalfont and \bf 1I \normalfont lead 
to inequivalent theories because ${\cal H}$ and ${\cal H}_i$ propagate independently.  
So starting from some constraint and the demand of integrability might miss out 
independent but compatible constraints.  \bf 1 \normalfont and \bf 1I \normalfont are 
however equivalent (by inspection of the constraint algebras)
for GR coupled to the known fundamental bosonic fields.

So far, at least the bosonic sector of nature appears to be much simpler than the GC  
hypersurface framework of GR might suggest, and the 3-space approach may be formulated
in two equivalent ways \bf 1 \normalfont and \bf 1I \normalfont as regards best matching.  
We now consider both \bf 1 \normalfont and \bf 1I \normalfont for spin-$\frac{1}{2}$ fermions.

\subsection{Fermions and the 3-Space Approach}

Whereas it is true that the spinorial
laws of physics may be rewritten in terms of tensors \cite{PenroseRindler},
the resulting equations are complicated and it is not clear if and how they
may be obtained from action principles.  Thus we are almost certainly
compelled to investigate coupled spinorial and gravitational fields by attaching local flat frames to our manifolds.

There are two features we require for the analysis of the spin-$\frac{1}{2}$ laws of nature coupled to gravity.  
First, we want the analysis to be clear in terms of shift and lapse kinematics, given our success in this paper with this approach.  
However, one should expect the spinors to have further sorts of kinematics not present for tensor fields.  
Second, we want to explicitly build SO(3,1) (spacetime) spinors out of SO(3) (spatial) ones.\fn{This is 
standard use of representation theory, based on the accidental 
Lie algebra relation $SO(4)\cong SO(3) \bigoplus SO(3)$, 
which depends on the dimension of space being 3.  
This relation is a common source of tricks in the particle physics and quantum gravity literatures.  
By SO(3,1) and SO(3) spinors, we strictly mean spinors corresponding to their universal covering groups, SL(2,C) and SU(2) respectively.  
We are not yet concerned in this paper with the differences betwen SO(4) and SO(3,1) from a quantization perspective, which render Euclidean quantum programs easier 
in some respects.}  We hope to perform this first-principles analysis in the future.  In this paper, we consider the first feature in the following 4-component spinor 
formalism.    

In G\'{e}h\'{e}niau and Henneaux's (GH) \cite{GH} 4-component spinor study of the Einstein--Dirac (ED) system, 
the term $\bar{\psi}\gamma^{\bar{\lambda}}\nabla^{\mbox{\scriptsize s\normalsize}}_{\bar{\lambda}}\psi$ is decomposed as follows\fn{ We use 
barred Greeks for Minkowski indices and 
barred Latins for Euclidean indices. 
The Minkowski metric is denoted by $\eta_{\bar{\mu}\bar{\nu}}$. 
The $\gamma^{\bar{\lambda}}$ are Dirac matrices, 
obeying the Dirac algebra $\gamma^{\bar{\mu}}\gamma^{\bar{\nu}} + \gamma^{\bar{\nu}}\gamma^{\bar{\mu}} = 2\eta^{\bar{\mu}\bar{\nu}}$, 
which is not to be confused with the Dirac Algebra (\ref{Algebra}).     
Dirac's suited triads are denoted by $E_{\bar{\lambda}}^{\mu}$; 
these obey $E_{\bar{0}a} = 0$, 
$E_{\bar{0}0} = - N$, 
$E_{\bar{\mu}}^{\sigma} E_{\bar{\nu}\sigma} = \eta_{\bar{\mu}\bar{\nu}}$ and 
$E_{\bar{\lambda}\alpha} E^{\bar{\lambda}}_{\beta} = g_{\alpha\beta}$.  
$\psi$ is a 4-component spinor, with conjugate $\bar{\psi} = i\gamma^{\bar{0}}\psi^{\dagger}$.  
The spacetime spinorial covariant derivative is 
$\nabla^{\mbox{\scriptsize s\normalsize}}_{\bar{\nu}}\psi = \psi_{,\bar{\nu}} - \frac{1}{4}\Omega_{\bar{\rho}\bar{\sigma}\bar{\nu}}\gamma^{\bar{\rho}}\gamma^{\bar{\sigma}}\psi$, 
where $\Omega_{\bar{\rho}\bar{\sigma}\bar{\nu}} = (\nabla_{\beta}E_{\bar{\rho}\alpha})E^{\alpha}_{\bar{\sigma}} E^{\beta}_{\bar{\nu}}$ is the spacetime spin connection.  
The spatial spinorial covariant derivative is 
$D^{\mbox{\scriptsize s\normalsize}}_{\bar{p}}\psi = \psi_{,\bar{p}} - \frac{1}{4}\omega_{\bar{r}\bar{s}\bar{p}}\gamma^{\bar{r}}\gamma^{\bar{s}}\psi$, 
where $\omega_{\bar{r}\bar{s}\bar{p}} = (D_{b}E_{\bar{r}a})E^{a}_{\bar{s}} E^{b}_{\bar{p}}$ is the spatial spin connection.  }
$$
\sqrt{|g|}\bar{\psi}\gamma^{\bar{\lambda}}\nabla^{\mbox{\scriptsize s\normalsize}}_{\bar{\lambda}}\psi = i\sqrt{h}{\psi}^{\dagger}
\left[
N\gamma^{\bar{0}}\gamma^{\bar{l}}D^{\mbox{\scriptsize s\normalsize}}_{\bar{l}}\psi + \frac{NK}{2}\psi 
\right.
$$
\be
\left.
+ N_{,\bar{l}}\gamma^{\bar{0}}\gamma^{\bar{l}}\psi  
- (\dot{\psi} - \pounds^{\mbox{\scriptsize s\normalsize}}_{\xi}\psi - \pa_{\mbox{\scriptsize R\normalsize}}\psi)
\right],  
\label{spinsplit}
\ee
where 
\be
\pounds^{\mbox{\scriptsize s\normalsize}}_{\xi}\psi = \xi^i\psi_{,i} - \frac{1}{4}
E_{[\bar{r}|}^i\pounds_\xi E_{|\bar{s}]i}\gamma^{\bar{s}}\gamma^{\bar{r}}\psi,
\label{splied}
\ee
\be
\pa_{\mbox{\scriptsize R\normalsize}}\psi = \frac{1}{4}E^i_{[\bar{r}}\dot{E}_{\bar{s}]i}\gamma^{\bar{r}}\gamma^{\bar{s}}\psi.
\label{Rottcod}
\ee
First, observe that the tensorial Lie derivative $\pounds_{\xi}\psi = \xi^i\psi_{,i}$ 
is but a piece of the spinorial Lie derivative (\ref{splied}) \cite{GH, Kosmann}.  
There is also an additional triad rotation correction (\ref{Rottcod}) to the velocities 
in addition to the 3-diffeomorphism-dragging Lie derivative correction.    
The notion \bf 1 \normalfont of best matching must be generalized to accommodate this additional, 
very natural geometric correction: given two spinor-bundle 
3-geometries $\Sigma_1$, $\Sigma_2$, the (full spinorial) drag shufflings of $\Sigma_2$ 
(keeping $\Sigma_1$ fixed) are accompanied by the rotation shufflings of the triads glued to it.      
The triad rotation correction is associated with a further `locally Lorentz' constraint 
${\cal J}_{\bar{\mu}\bar{\nu}}$ \cite{Diracrec}.  

In thinking from first principles about best matching in sufficiently general terms 
to include the treatment of spinors, it is not clear whether the triad rotations 
need be included from the start.  One might `discover and encode' these as occurs 
with the Gauss laws for 1-forms.  Also, use of the `bare' principle \bf 1I \normalfont 
may not require a conceptual advance on best matching:  the Dirac procedure beginning 
with ${\cal H}$ would provide us with the correct ${\cal H}_i$, whose encoding would 
yield the full $\xi^i$ correction for spinors.  Pursuing this last line of approach, 
Nelson and Teitelboim's work \cite{fermi} may be taken to imply that 
${\cal H}_i$ and ${\cal J}_{\bar{\mu}\bar{\nu}}$ are indeed integrability conditions for ${\cal H}$.  
For in terms of Dirac brackets $\{\mbox{ }, \}^*$, starting from ${\cal H}$, $\{{\cal H}, {\cal H}\}^*$
gives ${\cal H}_i$ and then we can form $\{{\cal H}, {\cal H}_i\}^*$ which
gives ${\cal J}_{\bar{\mu}\bar{\nu}}$ (and ${\cal H}$) so we have recovered all the constraints as
integrability conditions for ${\cal H}$.  
One does not recover ${\cal H}$ if one starts with ${\cal H}_i$ or ${\cal J}_{\bar{\mu}\bar{\nu}}$, so in some sense ${\cal H}$ is privileged.  
However, this does highlight our other caveat for the integrability idea: one might choose to 
represent the constraint algebra differently by mixing up the usual 
generators.  For example, a linearly-related set of 
constraints is considered in \cite{fermi}, for which the integrability of any of the 
constraints forces the presence of all the others.    
Our defence against this is to invoke again that we only require one formulation of the 3-space approach to work, so we would begin with the 
quadratic constraint ${\cal H}$ nicely isolated.

Second, although derivative coupling (second term) and tilt (third term) appear to be present in (\ref{spinsplit}), GH observed that these cancel in the Dirac field  
contribution to the Lagrangian density,

\be
\sqrt{|g|}\mbox{\sffamily L\normalfont}_{\mbox{\scriptsize D\normalsize}} 
= \sqrt{|g|}\left[\frac{1}{2}(\bar{\psi}\gamma^{\bar{\lambda}}\nabla^{\mbox{\scriptsize s\normalsize}}_{\bar{\lambda}}\psi
- \nabla^{\mbox{\scriptsize s\normalsize}}_{\bar{\lambda}}\bar{\psi}\gamma^{\bar{\lambda}}\psi) - m_{\psi}\bar{\psi}\psi\right]
\label{fermilag}.
\ee
Whilst Nelson and Teitelboim \cite{fermi} do not regard their formulation's 
choice of absence of derivative coupling as a deep simplification (they adhere to the HKT school of thought 
and the simplification is not in line with the hypersurface deformation algebra), the GH result is clearly encouraging for the 3-space approach.  
For, once (\ref{spinsplit}) has been used in (\ref{fermilag}), 
we obtain an action of the form \bf 2UG\normalfont, so we can cast ED theory into the \bf 2G \normalfont generalized BSW form (\ref{actualfermi}).  

Finally, we comment on the inclusion of 1-form--fermion interaction terms of the Einstein--Standard Model theory
\be
i\mbox{\sffamily f\normalfont}_{\cal A}\tau^{\cal A}_{\mbox{\scriptsize\bf I\normalfont\normalsize}}\bar{\psi}\gamma^{\bar{\beta}}
E^{\mu}_{\bar{\beta}}A^{\mbox{\scriptsize\bf I\normalfont\normalsize}}_{\mu}\psi
\label{star}
\ee
where ${\cal A}$ takes the values U(1), SU(2) and SU(3) and $\tau^{\cal A}_{\mbox{\scriptsize\bf I\normalfont\normalsize}}$ are the generators of these groups. 
The decomposition of these into spatial quantities is trivial.  
No additional complications are expected from the inclusion of such terms, since  
1) they contain no velocities so the definitions of the momenta are unaffected (this includes there being no scope for derivative coupling) 
2) they are part of gauge-invariant combinations, unlike the Proca term which breaks gauge invariance and significantly alters the Maxwell canonical theory.  
In particular, the new terms clearly contribute linearly in $A_{\perp}$ to the Lagrangian potential, so by the argument at the end of Sec IV.A, an accident 
occurs ensuring that tilt kinematics is not necessary.  Also, clearly the use of the form (\ref{fermilag}) is compatible with the inclusion of the interactions (\ref{star}) 
since, acting on $\bar{\psi}$ the gauge correction is the opposite sign.  So our proposed formulation's combined Standard Model matter Lagrangian is  
$$
\mbox{\sffamily L\normalfont}_{\mbox{\scriptsize SM\normalsize}}^{\cal A} = 
\left[
\frac{1}{2}(\bar{\psi}\gamma^{\bar{\lambda}}
(\nabla^{\mbox{\scriptsize s\normalsize}}_{\bar{\lambda}} 
- i\mbox{\sffamily f\normalfont}_{\cal A}\tau^{\cal A}_{\mbox{\scriptsize\bf I\normalfont\normalsize}}
E^{\mu}_{\bar{\lambda}} A^{\mbox{\scriptsize\bf I\normalfont\normalsize}}_{\mu})\psi
\right.
$$
\be
\left.
- (\nabla^{\mbox{\scriptsize s\normalsize}}_{\bar{\lambda}} 
+ i\mbox{\sffamily f\normalfont}_{\cal A}\tau^{\cal A}_{\mbox{\scriptsize\bf I\normalfont\normalsize}}
E^{\mu}_{\bar{\lambda}}A^{\mbox{\scriptsize\bf I\normalfont\normalsize}}_{\mu})     )\bar{\psi}\gamma^{\bar{\lambda}}\psi) - m_{\psi}\bar{\psi}\psi\right] 
+ \mbox{\sffamily L\normalfont}_{\mbox{\scriptsize YM\normalsize}}^{{\cal A}}.  
\ee
Here $\mbox{\sffamily L\normalfont}_{\mbox{\scriptsize YM\normalsize}}^{\cal{A}}$
is given by the $m = 0$ version of (\ref{lagYMmass}) and we would need to sum the 
square bracket over all the known fundamental fermionic species,
which thus simultaneously incorporates all the required accidents.    
There is also no trouble with the incorporation of the Yukawa interaction term $\bar{\psi}\chi\psi$ which could be 
required for some fermions to pick up mass from a Higgs scalar.  

Thus the Lagrangian for all the known fundamental matter fields can be built by assuming best-matching kinematics and that the DeWitt structure 
is respected.  The thin sandwich conjecture can be posed for all these fields coupled to GR.  The classical physics of all these fields is timeless in Barbour's sense.

\subsection{Future Developments}

We end by suggesting further work toward answering Wheeler's question in the introduction stimulated by the advances in this paper.    
It remains to explicitly build a best-matched generalized BSW ED action starting from a pair of spatial SO(3) spinors.  
Use of (\ref{spinsplit}) in (\ref{fermilag}) still has remnants of 4-dimensionality in its appearance: it is in terms of 4-component spinors and Dirac matrices.  
However, recall that the Dirac matrices are built out of the Pauli matrices associated with SO(3), and choosing to work in the chiral representation, the 4-component spinors 
may be treated as $\psi = [\psi_{\mbox{\scriptsize D\normalfont}}, \mbox{ } \psi_{\mbox{\scriptsize L\normalfont}}]$ 
i.e in terms of right-handed and left-handed SO(3) 2-component spinors.  
Thus a natural formulation of ED theory in terms of 3-dimensional objects exists.  
To accommodate neutrino (Weyl) fields, one would consider a single SO(3) spinor, that is set 
$\psi = [0, \mbox{ } \psi_{\mbox{\scriptsize L\normalfont}}]$, $m = 0$ before the variation is carried out.  
Whilst we are free to accommodate all the known fundamental fermionic fields in the 3-space approach, one cannot predict the number of Dirac and Weyl fields 
present in nature nor their masses nor the nongravitational forces felt by each field.  
So, consider actions with integrands such as
$\sqrt{R + U_{\mbox{\scriptsize F\normalsize}}}\sqrt{T_{\mbox{\scriptsize g\normalsize}}} 
+ T_{\mbox{\scriptsize F\normalsize}}$ or $N(R + U_{\mbox{\scriptsize F\normalsize}}) 
+ \frac{1}{4N}T_{\mbox{\scriptsize g\normalsize}} + T_{\mbox{\scriptsize F\normalsize}}$
for $U_{\mbox{\scriptsize F\normalsize}}$ and $T_{\mbox{\scriptsize F\normalsize}}$ \it built from spatial first principles using SO(3) spinors \normalfont.  
Obtain ${\cal H}$ and treat its propagation exhaustively to obtain constraint algebras.
Is a universal light-cone recovered?
Is Einstein--Dirac theory singled out? 
One could attempt this work for a bare $T_{\mbox{\scriptsize F\normalsize}}$ 
or (more closely to BF\'{O}'s original work) for a best-matched $T_{\mbox{\scriptsize F\normalsize}}$. 
In connection with the latter, how is the thin sandwich conjecture for Einstein--Dirac theory well-behaved?    
On coupling a 1-form field, do these results hold for Einstein--Maxwell--Dirac theory?
On coupling $K$ 1-form fields, do they hold for Einstein--Yang--Mills--Dirac 
theories such as the Einstein--Standard Model?  There is also the issue of whether conformal gravity can accommodate 
spin-$\frac{1}{2}$ fermions.    

It is worth considering whether any of our ideas for generalizing the
3-space approach extend to canonical supergravity \cite{sugy}.  
This could be seen as a robustness test for our ideas and possibly lead to a new formulation of supergravity.  
Also, supersymmetry is proposed to resolve the hierarchy problem and help with many other problems of theoretical physics.  
Furthermore, if the hierarchy problem is to be resolved in this way, the forthcoming generation of particle accelerators are predicted to see superparticles.     
Hence there is another reason for asking if the 3-space approach extends to supergravity with supersymmetric matter: this may well be soon required to describe nature.    
The supergravity constraint algebra is not known well enough \cite{bookofdeath} to comment
whether the new supersymmetric constraint ${\cal S}_{\bar{\mu}}$ arises as an
integrability condition for ${\cal H}$.  Note however that Teitelboim was 
able to treat ${\cal S}_{\bar{\mu}}$ as arising from the square root of ${\cal H}$
\cite{sqrtteitel}; however this means that the bracket of ${\cal S}_{\bar{\mu}}$ and its conjugate 
gives ${\cal H}$, so it is questionable whether the supergravity ${\cal H}$
retains all of the primary importance of the GR ${\cal H}$.

Finally, given the competition from \cite{BOF1} and this paper, it would be interesting to see whether any variant\fn{Kouletsis' recent work \cite{Kouletsis}, 
comparison with which we consider beyond the scope of this paper, is a variation on HKT's work using the generally-covariant history formalism \cite{gchf}.  This work 
does not explicitly mention spin-$\frac{1}{2}$ fermions either.} of HKT can 
be made to accommodate spin-$\frac{1}{2}$ fermions, and also to refine Teitelboim's GR-matter postulates to the level of  HKT's pure GR postulates.

\section*{Acknowledgements}

EA is supported by PPARC.
We would like to thank Julian Barbour, Brendan Foster, Malcolm MacCallum, 
Niall \'{O} Murchadha and Reza Tavakol for discussions,
and the anonymous referee, Harvey Brown, Stephen Davis, Domenico Giulini, Nikolaos Mavromatos, Alexander Polnarev and Marek Szydlowski for helpful comments.
Finally, we would like to thank the organizers, funders and co-participants 
of the Bad Honnef Physikzentrum  Quantum Gravity School, during which Sec I--III took shape.


\begin{thebibliography}{99}
\bibitem{bookofdeath}  The most recent confirmation that we are aware of that this work remains incomplete is on p. 96 of D'Eath's book in \cite{sugy}.
\end{thebibliography}
\end{document}